\documentclass[dvipsnames, 10pt]{IEEEtran}
\usepackage[dvipsnames]{xcolor}
\usepackage{amsmath,amsthm,amssymb}
\usepackage{remark}
\usepackage{boxedminipage}
\usepackage{pgfplotstable}
\usepackage{float}
\usepackage{url}
\def\be{\begin{eqnarray}}
\def\ee{\end{eqnarray}}
\def\bee{\begin{eqnarray*}}
\def\eee{\end{eqnarray*}}
   
\newcommand{\ba}{\begin{align}} 
\newcommand{\ea}{\end{align}}
\newcommand{\revonetwo}{\color{black}}
\newcommand{\revonethree}{\color{black}}
\newcommand{\revonefour}{\color{black}}
\newcommand{\revonefive}{\color{black}}
\newcommand{\revoneseven}{\color{black}}
\newcommand{\revoneeight}{\color{black}}
\newcommand{\revonenine}{\color{black}}
\newcommand{\revtwoseven}{\color{black}}

\newcommand{\revtwoten}{\color{black}}

\usepackage{amsfonts} 
\usepackage{amssymb}

\def\be{\begin{eqnarray}}
\def\ee{\end{eqnarray}}
\def\bee{\begin{eqnarray*}}
\def\eee{\end{eqnarray*}}

\newremark{remark}{Remark}

\newcommand{\rvline}{\hspace*{-\arraycolsep}\vline\hspace*{-\arraycolsep}}
\def\bmx{\begin{pmatrix}}
\def\emx{\end{pmatrix}}

\begin{document}

\title{On the Resilience of DAG-based Distributed Ledgers in IoT Applications}

\author{A. Cullen\thanks{Andrew Cullen, Pietro Ferraro and Robert Shorten are with the Dyson School of Design Engineering - Imperial College London.}, P.~Ferraro,
  C.~King,\thanks{Christopher King is with the Department of Mathematics, Northeastern University, Boston, MA 02115 USA.} and
  R.~Shorten}%
  
  \maketitle
  
\begin{abstract}
Distributed Ledgers have been proposed for a number of applications in the IoT domain where it is essential to have an immutable and irreversible record of transactions. Directed Acyclic Graph (DAG)-based architectures, in particular, seem  to provide a vast array of advantages over the more traditional Blockchain; however it can be challenging to conduct a thorough analysis of DAG-based ledgers and derive reliable performance
guarantees. In this paper, we analyse one commonly discussed attack scenario known as the \emph{parasite chain attack}, which aims at disrupting the immutability and irreversibility of the ledger, in the context of the IOTA Foundation's DAG-based system. Using a Markov chain model we study the vulnerabilities of IOTA's core Tip Selection method against this attack and we present an extension of the algorithm to improve the resilience of the ledger in this scenario.
\end{abstract}

\section{Introduction}

With an ever increasing number of connected devices---predicted to be 50 billion by 2022 \cite{Khan}---the Internet of Things (IoT) has become, in recent years, a topic of great interest in both academia and industry. The power of IoT comes from the interaction of numerous small and lightweight computing elements; this interaction is expected to provide access to a vast array of distributed services, leading to opportunities for innovation at all levels. From the point of view of a single agent, the introduction of IoT will affect both worklife and homelife: for example domotics \cite{Miori}, health-related applications \cite{Bhatt} and enhanced learning \cite{Farhan} are areas where these changes will impact daily life.
At the same time, from a broader point of view, IoT is expected to bring advantages to the community as a whole, by providing services such as smart mobility \cite{oldpaper}, detecting weather conditions \cite{Li} and monitoring surgery in hospitals \cite{Yeole}. However several challenges still need to be addressed in order to turn this vision into a reality. First of all, the constrained capabilities of many IoT devices as well as the architectures of current access control systems, which are based on centralized and hierarchical structures, impede the interaction of multiple IoT machines (which are often not in the same trust domain)  \cite{Dorri}. Secondly, the kind of data that is shared amongst IoT agents can be privacy-sensitive or safety-critical and hence an appealing target for various cyber attacks. Devoting precious energy and computational resources to support security is often infeasible due to CPU, memory and battery constraints \cite{Dorri} \cite{Mahmoud}.\newline

Concurrent with this growth of interest in IoT, Distributed Ledger Technologies (DLTs)---the agnostic term for Blockchain and related technologies--- have emerged as an attractive solution for the problem of distributed consensus in a database \cite{Dorri}. In the aftermath of the 2008 financial crisis, Satoshi Nakamoto (a pseudonym for the original author or group of authors) proposed in their whitepaper \cite{Nakamoto} the use of an architecture, called the Blockchain, to solve the problem of distributed consensus on a ledger. Since then, the scientific community has been probing the boundaries of this new technology with the goal of applying it beyond financial transactions. More specifically DLTs have recently emerged as an enabling technology for managing, controlling and securing interactions in a range of cyberphysical systems \cite{Samaniego}\cite{Novo}\cite{Panarello}. We will argue that due to their distributed and trustless nature, DLTs have characteristics that would prove advantageous as a data transfer and transaction settlement layer for the IoT domain:

\begin{itemize}
\item{\em Decentralised Trust:}  DLTs introduce a distributed consensus mechanism whereby users in a P2P network can interact and exchange data without the need for a trusted third party to guarantee the integrity of the ledger. In IoT this would allow devices with mutually exclusive trust domains to interact with one another without the need of a centralised server;\newline
\item{\em Irreversibility, Immutability and Transparency:} {\color{black}A DLT grants an irreversible, immutable and transparent record of transactions}. Each agent owns a copy of the ledger and can verify the correctness of any sequence of actions. In IoT this would allow multiple devices to orchestrate their behaviour in order to achieve a common goal;\newline
\item{\em Pseudo-Anonymity and Data Sovereignty:} Transactions in a DLT are pseudo-anonymous\protect\footnote{https://laurencetennant.com/papers/anonymity-iota.pdf}, due to the cryptographic nature of the addressing, and their contents can be encrypted. This allows each user to control access to and ownership of their own data;\newline
\item{\em Communication Security:} The cryptographic nature of DLTs means that problems such as key management and distribution are handled intrinsically. In IoT each device would have its own unique ID and asymmetric key pair, and so light-weight security protocols (that would fit and stratify the requirements for the limited resources of IoT devices) become more feasible.
\end{itemize}

There are many varieties of DLT architecture, but not all are equally useful for IoT applications. In particular Blockchain, while responsible for the initial explosion of interest in DLT, has limited use in IoT applications: the large energy cost of mining, the volatility of bitcoin, long transaction approval times, transaction fees, and the inherent preference of miners to process large transactions rather than small ones \cite{oldpaper}, all create a bottleneck when thousands of devices communicate with each other many times per second. To overcome these problems, several solutions have been explored: in \cite{Dorri2} the authors describe a Lightweight Scalable Blockchain (LSB) that is optimized for IoT requirements; \cite{Novo} proposes the use of management hub nodes to overcome scalability and computing constraints; while in \cite{Dorri} the authors tackle Blockchain limitations with an always online, high resource device known as miner that is responsible for handling all communication
within and external to the home. In this paper we focus on a different architecture for DLTs based on Directed Acyclic Graphs (DAGs), and more specifically, on one DAG-based ledger known as the IOTA Tangle \cite{Popov}. This architecture seems to possess properties that make it suitable for the IoT domain; for a more thorough comparison between Blockchain-based DLTs and DAG-based DLTs, in the IoT context, we refer the interested reader to \cite{oldpaper}. In particular, the aim of this paper is to explore the security properties of the IOTA Tangle: as mentioned earlier, IoT devices often share privacy-sensitive or safety-critical data, and so the security properties of the ledger are essential for its success in IoT applications. Since the resilience of DLTs to cyber attacks is of paramount interest to the IoT domain, in this paper we analyse the response of the IOTA Tangle to a commonly discussed attack scenario known as a \emph{parasite chain attack}, which is akin to \emph{long range attacks} in blockchains and is considered to be one of most effective attack vector for doublespending on a DAG-based DLT (the term `doublespending' here refers to an attempt to alter or reverse any record previously added to the ledger). {\color{black} The safety of the system is determined by an algorithm, called Biased Random Walk (BRW) (thoroughly analysed in the reminder of this paper), whose objective is to leverage the total computational power of the honest users of the network to maintain the ledger safe against such attacks. Unfortunately, due to the complexity of the system at hand tuning the algorithm's parameters and its security against double spending attacks results in a non trivial task. Therefore, to circumvent this issue and increase the overall security of the ledger, in this paper we present an extension of the BRW algorithm and we validate its efficacy through simulations}.\newline
To summarize, the contributions of this paper are:

\begin{itemize}
	\item A mathematical description of the DAG-based DLT known as the Tangle, together with a Markov chain model for the algorithm controlling how transactions are added;
	\item A thorough analysis of the \emph{parasite chain attack} on the Tangle, wherein the attacker attempts to alter or reverse one or more transactions previously added to the ledger;
	\item A proposal for a new extension of the BRW algorithm, which is shown to improve the resistance of the ledger to parasite chain attacks.\newline
\end{itemize} 
 
The remainder of this paper is organised as follows: in Section \ref{sec: Tangle} we describe the Tangle DAG in detail and introduce a double spending mechanism known as a parasite chain attack. Section \ref{sec: BRW} summarizes the stochastic model for the Tangle and presents a new  formulation for the Biased Random Walk (BRW) tip selection algorithm as an absorbing Markov chain. This formulation is used to analyse  the algorithm's resistance to parasite chain attacks over a range of parameter choices. Section \ref{sec: first order BRW} introduces a modification to the BRW which makes use of the growth of the cumulative weight in the Tangle and presents  results showing its resistance to a parasite chain attack.

\section{The Tangle}
\label{sec: Tangle}
\begin{figure}
\centering
\includegraphics[width=1\columnwidth]{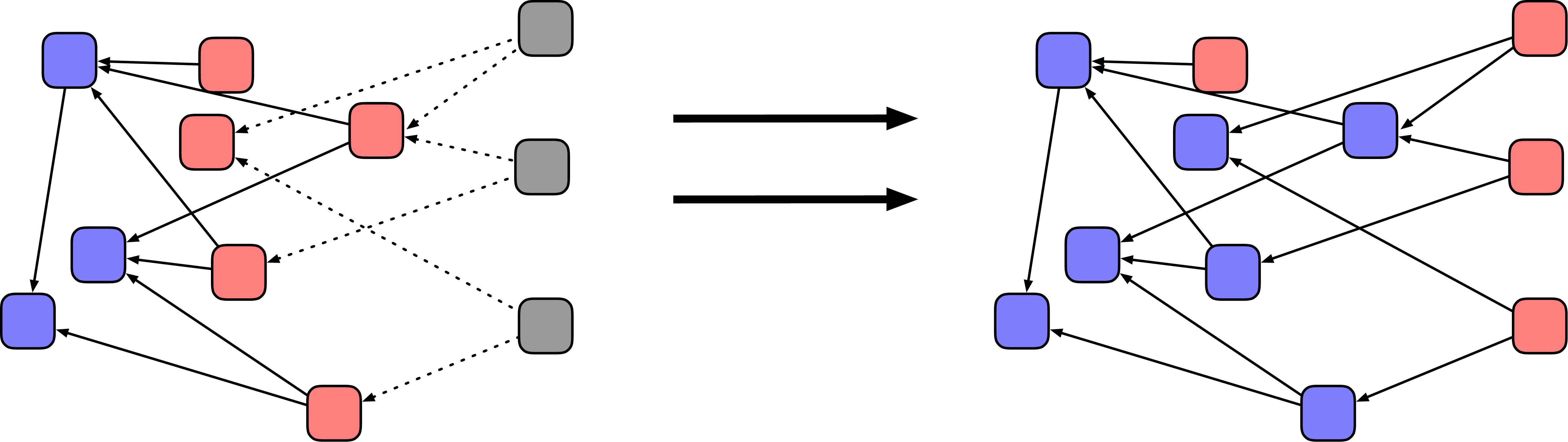}
\caption{Sequence to issue a new transaction. {\revtwoten The blue sites represent transactions that have received at least one direct approval}, the red ones represent the tips and the gray ones represent newly arriving transactions. The black edges represent approvals, whereas the dashed ones represent transactions that are performing the PoW in order to approve two tips. After completing the PoW approved transactions cease to be tips.}
\label{Fig: Tangle}
\end{figure}

\begin{figure}
\includegraphics[width=1\columnwidth]{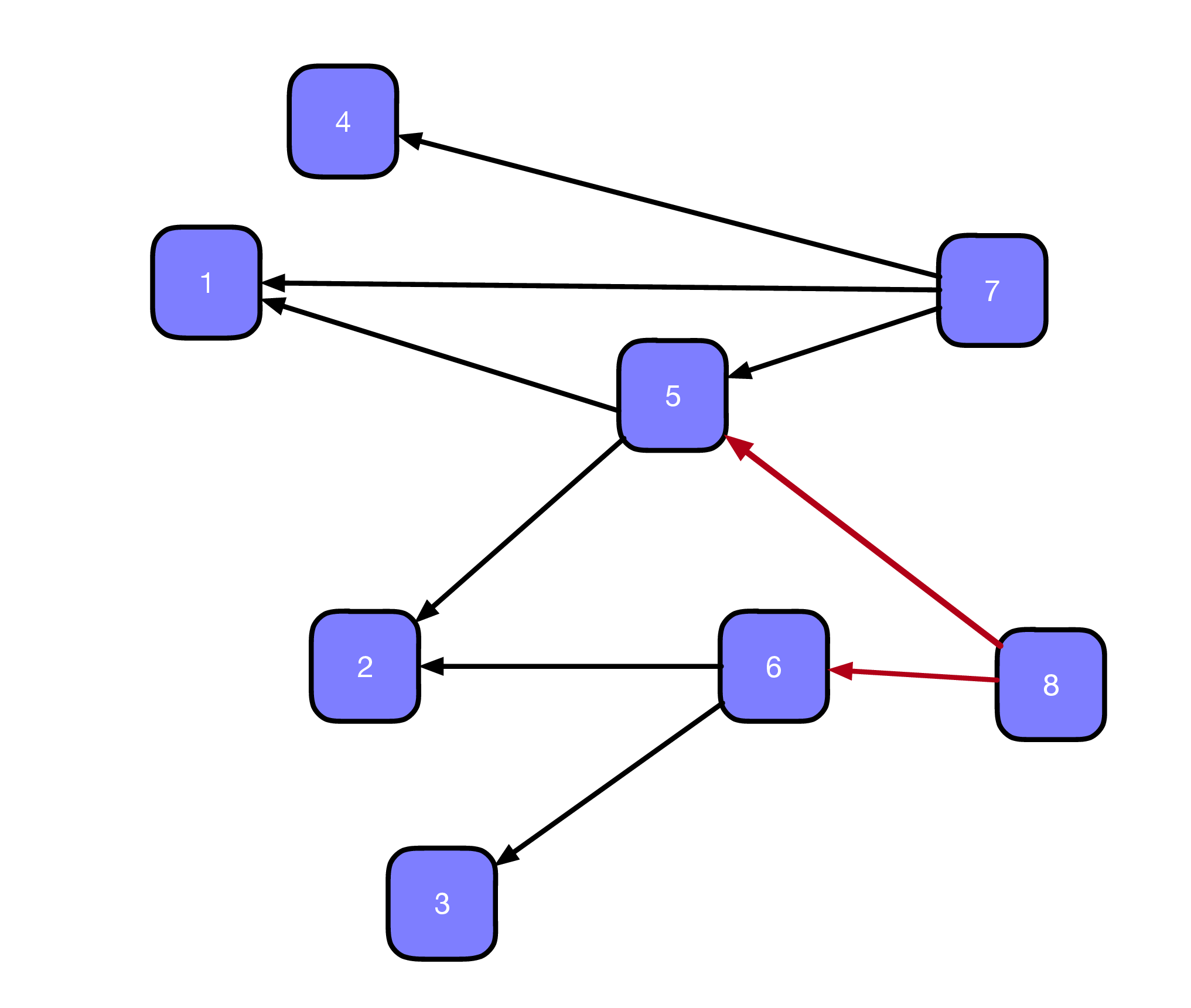}
\caption{Transaction 8 directly approves 5 and 6. It indirectly approves 1, 2 and 3. It does not approve 4 and 7. This image was also present in \cite{oldpaper}.}
\label{Fig: DirUndir Approval}
\end{figure}

The Tangle is a distributed ledger where transactions are stored in a Directed Acyclic Graph (DAG, i.e. a finite directed graph without any cycles) and where every new transaction validates several previous transactions using a Proof of Work (PoW) mechanism\footnote{Recently, the IOTA foundation has considered replacing the
PoW based mechanism with one based on Verifiable Delay Functions (VDFs) \cite{Boneh}}. New transactions are added to the Tangle by issuers known as \emph{nodes}, {\color{black}that unlike the miners in
Blockchain  \cite{Nakamoto}, do not receive any payment for their work}. Whenever a new transaction (in what follows, we will use transaction, site and vertex interchangeably) is issued, $m$ previous transactions (normally set to $m=2$)  are selected and, after a small PoW, a site for the new transaction and $m$ new edges  are added to the Tangle {\revtwoten (by selecting a site, a node is approving this site as free from conflicts etc.)}; see Figure \ref{Fig: Tangle} for a visual representation of this process. {\revonetwo The amount of work spent to issue each transaction determines its own \emph{weight} \cite{Popov}; {\color{black} therefore a user might employ more hashing power, spending more time or more computational effort in performing the PoW, to give his/her transaction a larger weight}. In what follows, to increase the readability of the exposition, we assume that the weight of each transaction is fixed to 1. Note that this assumption does not affect the main results of the paper and can be easily relaxed.}\newline
Whenever a new edge is attached to site $j$ by the newly added site $i$ we say that $j$ is directly approved by $i$. If there is a directed path from $i$ to $j$, with length greater than one, we say that $j$ is indirectly approved by $i$ (e.g.,  see Figure \ref{Fig: DirUndir Approval}).
A transaction that has received no approvals is called a \emph{tip}, and the set of all these transactions is called the \emph{tips set}. In the IOTA protocol the node selects $m$ sites for approval from the current tips set; we discuss shortly several possible algorithms for the tip selection. 
 

\begin{figure}
\includegraphics[width=1\columnwidth]{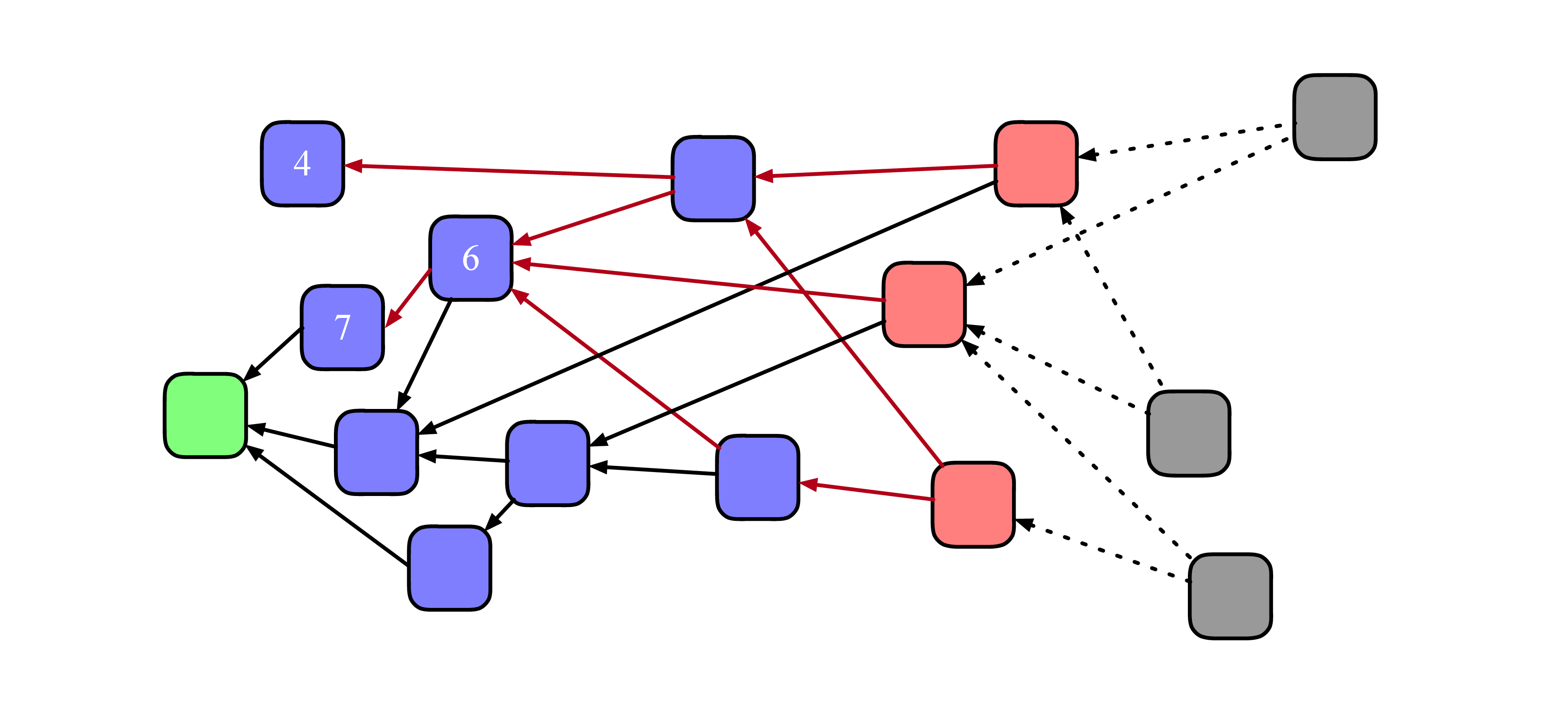}
\includegraphics[width=1\columnwidth]{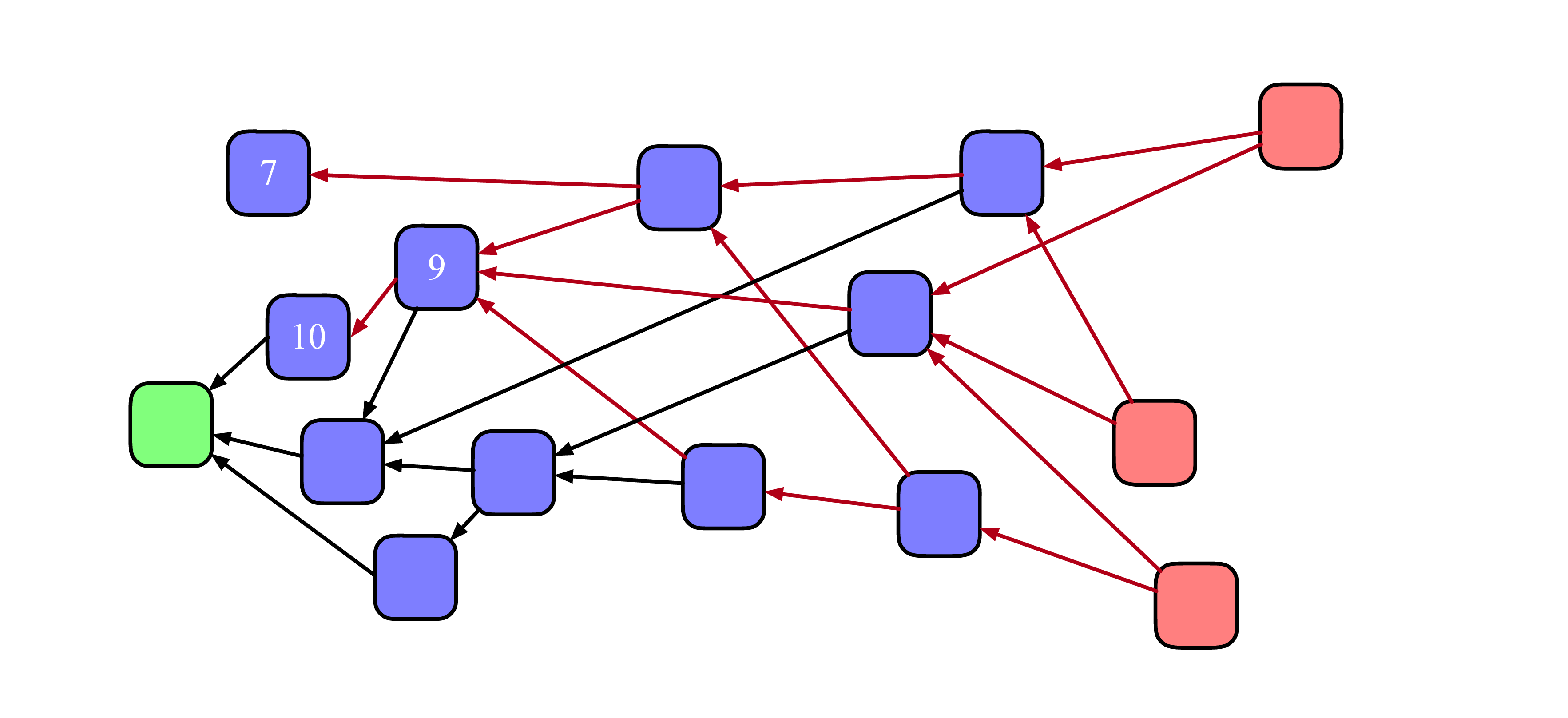}
\caption{Representation of the evolution of the Cumulative weight of three sites as three new transactions enter the Tangle}
\label{Fig: Cumulative weight}
\end{figure}
The concept of direct and indirect approval is crucial in our analysis. The core metric we consider to assess the security of the Tangle against attacks is the \emph{cumulative weight} $\mathcal{H}(t)$ of a transaction: {\revonetwo this value represents the weight of the transaction plus the number of vertices that, either directly or indirectly,  approve a given site. Roughly speaking, the cumulative weight can be interpreted as a relative measure of how legitimate a transaction is (i.e., a transaction with a higher cumulative weight is considered more trustworthy than a transaction with a lower one).} Figure \ref{Fig: Cumulative weight} shows an example of how the cumulative weight changes in time.  
We also introduce the idea of mutual consistency between transactions: two transactions are mutually consistent if there are no conflicts between the data contained in them or in all sites directly or indirectly approved by them. Notice that the idea of consistency is very general and is linked to the type of data that is being recorded on the ledger: for example it could refer to a conflict in the way a certain amount of currency was spent, or to a conflict in the records stored by two devices.
In what follows, we assume that there is a simple and fast way, called the \emph{verification step}, which occurs during the approval step, to verify whether or not the sites selected for approval by a node are mutually consistent with each other. If the verification step detects an inconsistency, the tip selection process must be re-run until a set of consistent sites is found. 

\begin{figure}
\includegraphics[width=1\columnwidth]{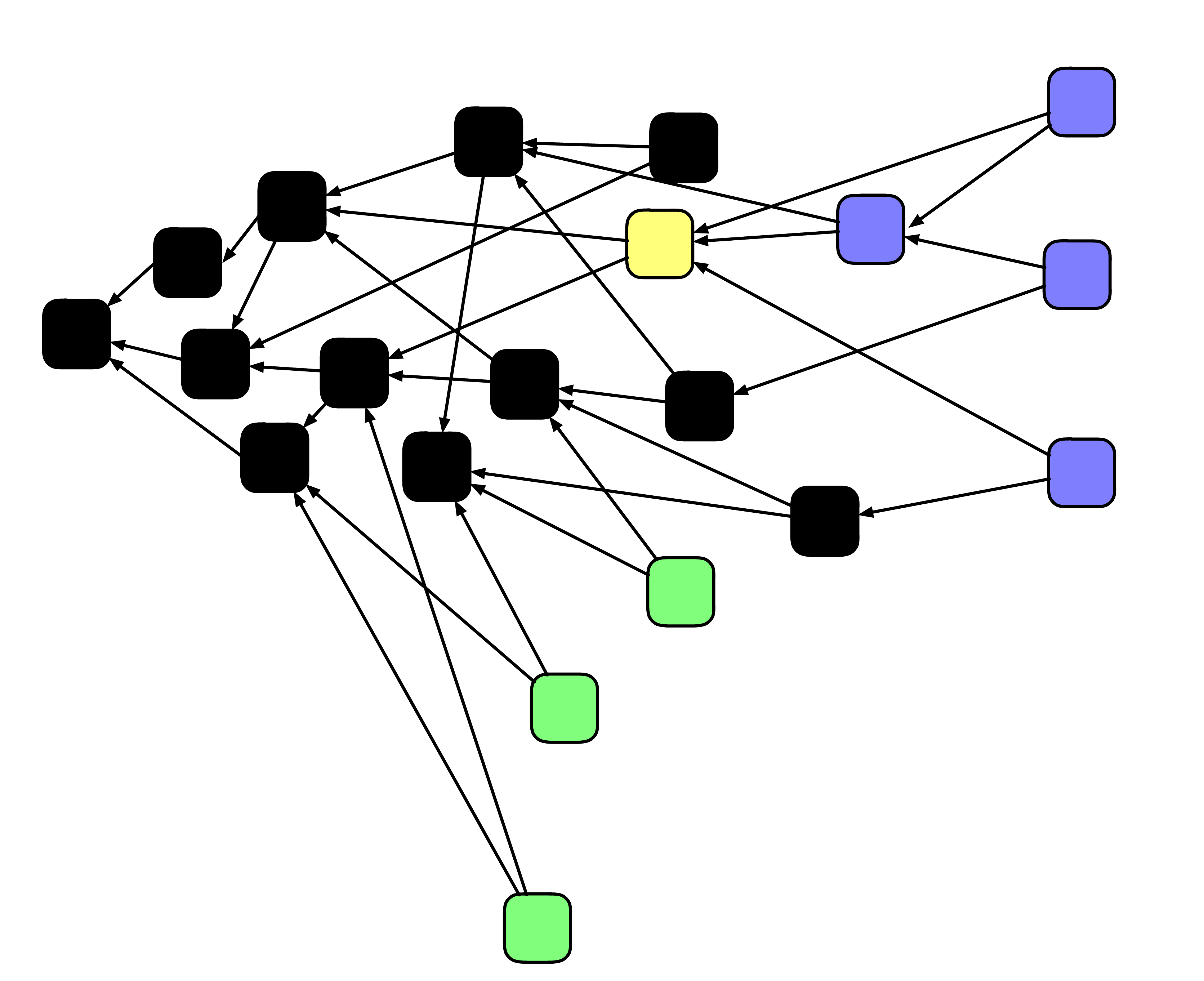}

\caption{The blue and the green transactions are incompatible with each other.  This image was also present in \cite{oldpaper}.}
\label{Fig: DoubleSpending}
\end{figure}

Let us consider an example to develop this idea further: Figure \ref{Fig: DoubleSpending} shows an instance of the Tangle. A malicious user writes some data to the ledger, and the corresponding site is the yellow block in the figure. The same user, afterwards, writes an amount of conflicting data to the ledger, corresponding to the green blocks. It is worth stressing, at this point, that since the ledger is shared in a P2P network, there is no mechanism to force a user to select certain sites for approval. Any site can be selected as long as it is mutually consistent with the sites that are approved (directly or indirectly) by it. Nevertheless it is reasonable to assume that the vast majority of nodes would have little interest in approving specific sites and would follow the tips selection algorithm used by the protocol. In this scenario, all the transactions that approve the original yellow site (the blue blocks in Figure \ref{Fig: DoubleSpending}) are inconsistent with the green ones, and therefore any new sites can either approve the green/black sites or the blue/black ones. The green/blue combination would be considered invalid and a new selection would be made. The objective of an hypothetical attacker would be then to wait for the original data (the yellow block) to be acted upon in some way, and then to release the double spend sites (the green blocks) in such a way that they get approved by the majority of the network rather than the original data (thereby reversing the record of the data that has already been acted upon). \newline

\begin{figure*}[tb!]
\centering
\includegraphics[width=0.85\textwidth]{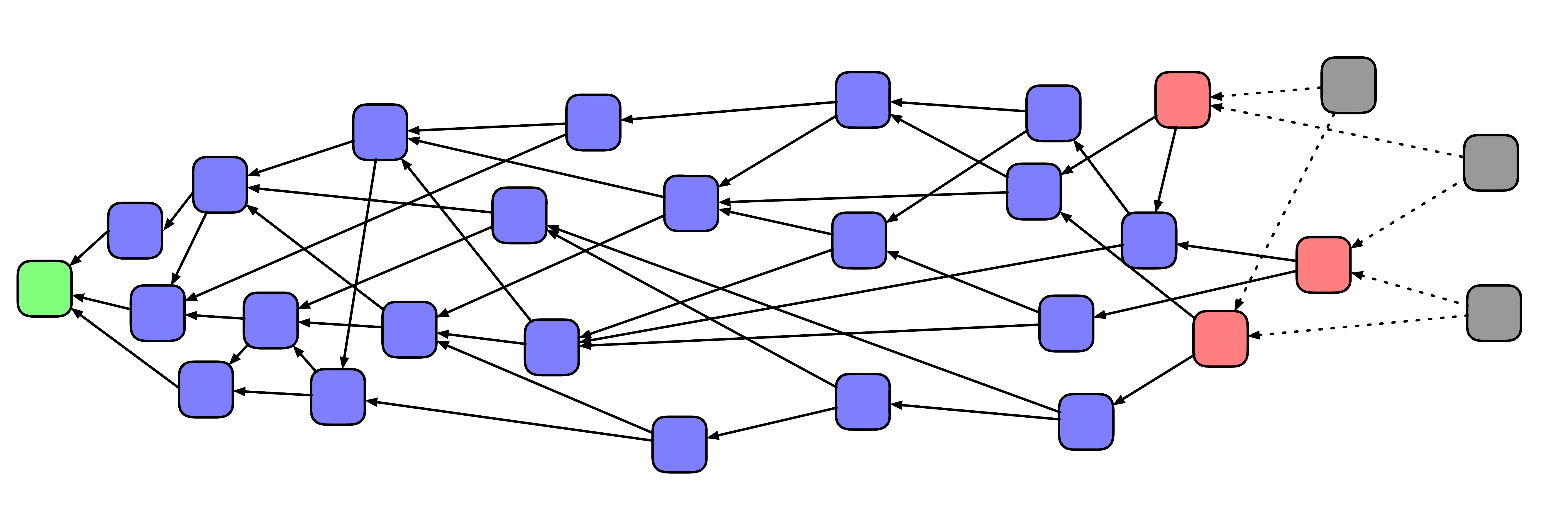}
\includegraphics[width=0.85\textwidth]{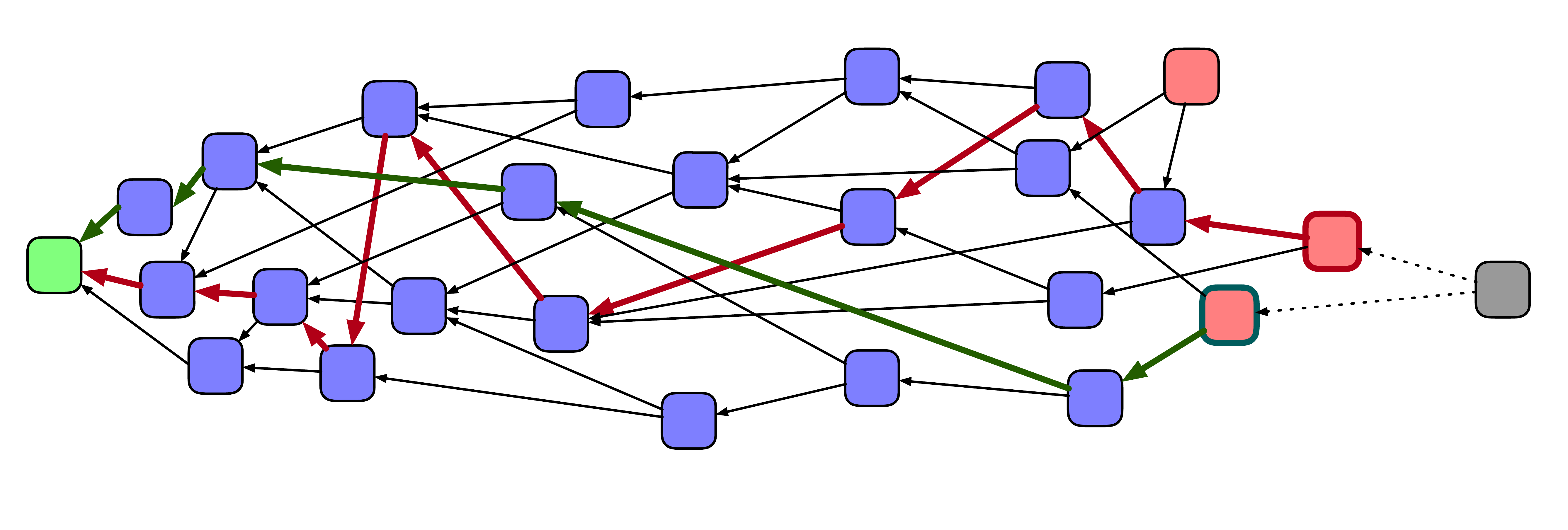}
\caption{Representation of the two main tips selection algorithms: the upper panel shows an instance of the Random Selection algorithm whereas the lower one presents a possible example of the Markov Chain Monte Carlo algorithm.}
\label{Fig: Selection Algorithms}
\end{figure*}

The success probability of such an attack depends on the way tips are selected by newly arrived transactions. Several tips selection algorithms have been proposed for the IOTA Tangle:\newline
\begin{itemize}
\item[1)] \emph{Uniform Random Tip Selection: } The Uniform Random Tip Selection (URTS) algorithm selects two tips randomly from the pool of all available tips. This algorithm, due to its simplicity, makes the Tangle vulnerable to double spending attacks. The upper panel of Figure \ref{Fig: Selection Algorithms} shows an example of the random selection procedure. The interested reader can refer to \cite{Popov} for a detailed discussion on this topic. \newline

\item[2)] \emph{Biased Random Walk  Algorithm: } The BRW algorithm (also referred to  as the Monte Carlo Markov Chain (MCMC) tip selection algorithm \cite{Popov}) is the main tip selection algorithm discussed in the original whitepaper of the IOTA Tangle \cite{Popov}. It works in a slightly more elaborate way than its random counterpart. In the BRW algorithm $m$ (generally two) independent random walks are created in the interior of the tangle; {\revtwoten the walks start at a site deep in the graph (ideally each walk would start from the genesis, but this quickly becomes infeasible as the graph grows. We discuss alternative start points in Section \ref{sec: BRW}) and move along the edges of the graph.} The probability of stepping along an edge from site $j$ to site $k$ is proportional to $f(- \alpha(\mathcal{H}_j-\mathcal{H}_k))$, where $f(\cdot)$ is a monotonic increasing function (generally an exponential), $\alpha$ is a positive constant and $\mathcal{H}_i$ represents the Cumulative Weight of site $i$. The stepping process stops when the particle reaches a tip, which is then selected for approval. The lower panel of Figure \ref{Fig: Selection Algorithms} shows an example of the paths of two walks in this selection procedure.

Ultimately the goal of the BRW algorithm is to increase the security of the ledger against double spending attacks performed by malicious users. 
The main difference between the BRW and URTS algorithms lies in the use of the graph structure: for BRW an attacker would need to create enough transactions, with cumulative weights equal to or larger than the cumulative weight of the main DAG, in order to make the double spending successful. This, in turn, would require the malicious user to possess an amount of computational power comparable to the network of honest users. The algorithm and its security from attacks can be tuned using the parameter $\alpha$: high values of $\alpha$ increase the probability that the particle will step to the transactions with the largest available cumulative weight (as $\alpha$ approaches $\infty$ the particles will move in a deterministic way), whereas lower values of $\alpha$, as the cumulative weights matter less and the stepping probability tends to become uniform as $\alpha$ approaches zero, make the algorithm and its output more unpredictable. A good way of picturing the effect of this parameter is by comparing it to the inverse temperature of a gas: the smaller the value of $\alpha$ the warmer the gas and vice-versa. The interested reader can find more details on this topic in {\revtwoten \cite{POBLB}}.  \newline 

\item[3)] \emph{Hybrid Selection Algorithm: } One of the drawbacks of the BRW is that as $\alpha$ increases, so also does the probability that a tip will not receive any approval by newly incoming transactions. Therefore a trade-off is required when tuning this value, between the security of the Tangle and its stability. To overcome this issue, the authors proposed in \cite{TACPaper} the hybrid selection algorithm.  Its aim is to provide a high level of security against attacks, while ensuring that all tips get eventually approved. This is done by making two different selections: a security selection where a high $\alpha$ BRW is used to ensure the security of the Tangle and a swipe selection where a low $\alpha$ or a URTS is used to ensure that eventually all transactions are selected.\newline

\end{itemize}

In what follows, we focus solely on the BRW algorithm. The interested reader can refer to \cite{oldpaper} and \cite{TACPaper} for a thorough analysis of the URTS and the hybrid selection algorithms.

\begin{figure}[h!]
\includegraphics[width=1\columnwidth]{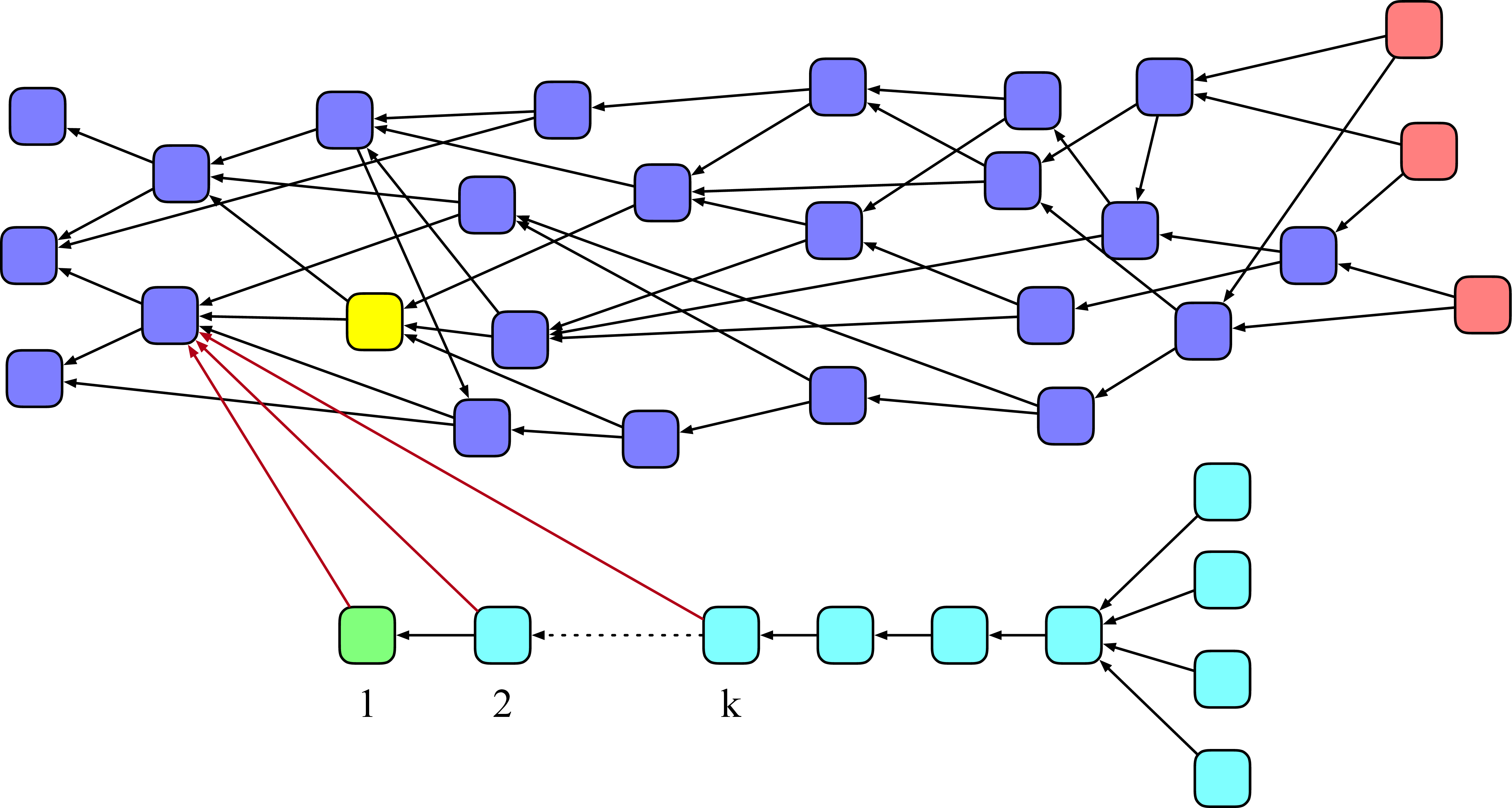}

\caption{A $k^{th}$-order SPC: The yellow and green transactions constitute a double spend.}
\label{Fig: kSPC}
\end{figure}
\subsection{The Parasite Chain Attack}
To see how in practice an hypothetical attacker could carry out a double spending in the Tangle (where the majority of users select tips using the BRW algorithm), we consider the attack scenario known as \emph{Parasite Chain Attack}. A Simple Parasite Chain (SPC) is illustrated in Figure \ref{Fig: kSPC}: we refer to this as $k^{th}$-order SPC because the first $k$ transactions in the chain reference the main Tangle. The attacker publishes the yellow transaction in the Tangle and simultaneously, in secret, creates a conflicting transaction (the green one) followed by a chain of vertices which validate it. The attacker waits for the yellow transaction to be confirmed\footnote{ It is generally accepted that a transaction should receive a minimum amount of approvals before it is considered to have been \emph{confirmed} (for instance in most Blockchain based cryptocurrencies, a transaction is considered valid after six subsequent approvals). Therefore the attacker should wait at least until the first transaction is confirmed and accepted by the Tangle before broadcasting the SPC.} and then immediately broadcasts the parasite chain to the network and continues publishing transactions which validate it. The goal of the attacker is then to try and ``race" the Tangle, creating a sub-DAG whose cumulative weight will outmatch the main body of the ledger. If the attack were to succeed, newly arrived transactions would attach to the branch that refers to the green transaction and would ultimately leave the honest site (the yellow one) behind.\newline

An SPC is characterised by four parameters:
\begin{itemize}
\item[1)] $T_{DS}$ is the time between the arrival of the original transaction and the broadcast of the SPC;
\item[2)] $k$ is the `order' of the SPC, i.e. the number of transactions referencing the main Tangle;
\item[3)] $\lambda$ is the rate of arrival of honest transactions i.e. the computing power of active honest network users
\item[4)] $\mu$ is the rate at which the attacker can add transactions to the parasite chain, i.e. the attacker's computing power (as opposed to $\lambda$ which represents the total computing power of honest users issuing transactions to the Tangle).
\end{itemize}

Note that it is possible to create parasite chains with more complex structure than the one presented previously. However due to the complexity of the analysis it is not yet clear how to optimally design such a chain.  Therefore we will focus solely on the SPC, and use this special case to draw conclusions about the Tangle's ability to resist doublespending attacks. {\color{black} A possible solution to this limitation, might be to employ machine learning algorithms (such as reinforcement learning), to guide the attacker's actions in designing the structure of the SPC to maximise his chance of success and, consequently, perform the same analysis that we propose in this paper on the shapes found this way. Nevertheless, this would be beyond the scope of this manuscript and will be investigated in a future work.}\newline  {\revonefour Finally, in the remainder of this paper we will assume that the \emph{honest majority assumption} holds true \cite{Popov}: the computing power of the attacker is always lower than the total computing power of the honest users accessing the network (i.e., $\mu<\lambda$).}\newline

\section{The Biased Random Walk Algorithm} \label{sec: BRW}
In this section we present a Markov Chain model for the BRW algorithm which allows us to compute the probability that a BRW will terminate on a given tip of a given instance of the Tangle. We refer to these probabilities as \emph{exit probabilities}. We then simulate a parasite chain attack on the Tangle, and investigate how the parameters of the BRW algorithm affect the Tangle's response to the attack. {\revoneeight In order to generate a random instance of the Tangle, in simulation, we use an agent based model in MATLAB: at each time step a random number of transactions arrive, according to a Poisson distribution, and for each one of these transactions, the tip selection algorithm is used to attach these to the Tangle.}
This agent based model simulates the arrival of each transaction, together with the tip selection procedure and the consistency checks, therefore providing an accurate replica of the mechanism previously described.

\subsection{Markov Chain Model}
{\revonefour We now describe the BRW as a Markov Chain process. Specifically, we explicitly formulate the transition matrix for the walk, which is an absorbing Markov chain: this allows us to obtain analytical expressions to compute the exit probabilities of the BRW.}
{ \revoneeight First we recall the description of the agent-based model which is used to generate a random instance of the Tangle.
Each new transaction selects $m=2$ tips for approval, and attempts to validate them.  To take into account possible conflicts we assume that $d$ conflicting sub-DAGs exist on the Tangle, where
each transaction from a sub-DAG is mutually consistent only with transactions from the same sub-DAG.
Thus every site has a label from the set $1,\dots,d$, indicating the sub-DAG to which it belongs. We will call this
the {\em type} of the site. If validation fails (i.e., transactions from different sub-DAGs are selected) the choices are discarded and another
two tips are selected for validation. This continues until the process is successful. {\color{black} Notice that, in a realistic scenario, the process of performing the BRW will take some time and it would be non realistic to model it as an instantaneous effort (especially in a IoT setting were computational power is already limited); nevertheless, since the modelling of this phenomena would not affect the the analysis on doublespending attacks (as it affects primarily the time evolution of the Tangle, by delaying the approval of all transactions), an analysis on the effects of the BRW's computational efforts are going to be the subject of a future work}.\newline 
 Once the validation process is over, there is a waiting period $h$ during which
the PoW is carried out.
During this time the approvals of the selected tips are pending, so the tips may still be
available for selection by other new transactions. After the waiting time $h$ the two sites which were successfully approved
are removed from the tips set, and so are no longer available for selection by other new transactions 
(at least, by the ones that follow the protocol)\footnote{It may happen that some of these sites had already ceased  to be tips at an earlier time, due to their being validated by some other new transaction.}. Finally, we assume that the arrival rate of newly issued transactions is determined by a Poisson distribution with rate $\lambda$.}

Next we review the BRW algorithm for tip selection: for a given instance of the Tangle, {\revoneseven a random walk is initiated starting somewhere in the interior of the graph. In our simulations, we use the genesis, but in practice, this start point needs to be selected by other means. In the current implementation of the Tangle, the walk begins from special sites known as milestones\footnote{https://blog.iota.org/coordinator-part-1-the-path-to-coordicide-ee4148a8db08} and exploring the way to select an optimal starting point remains an open question.} The BRW subsequently steps randomly along edges of the DAG. A step along an edge can be either forward (meaning toward a site that arrived later than the current site) or backward. A forward step occurs with a probability that is proportional to $\exp(- \alpha(\Delta \mathcal{H}))$, where $\alpha$ is a positive constant and $\Delta \mathcal{H}$ is the change in Cumulative Weight along the edge. The walk terminates when it reaches a tip, which is then selected for approval. Thus, to model the BRW algorithm  we need to define explicitly the cumulative weight of each transaction and find an expression for the probability of terminating on a given tip.

{\revonetwo Let $V(t)$ denote the set of sites on the DAG at time $t$. Since we set the weight of each transaction to 1, the cumulative weight $\mathcal{H}_i(t)$ of transaction $i$ is defined as one plus the number of transactions that directly or indirectly approve it at time $t$:
\be
\mathcal{H}_i(t)=1 + \#\lbrace z\in V(t) : z \  \mathrm{ approves } \  i  \rbrace.
\ee
}
The cumulative weight $\mathcal{H}_i(t)$ can be computed using the adjacency matrix $M(t)$ of the DAG ($[M]_{ij} = 1$ if there is a directed edge between transaction $j$ and transaction $i$, otherwise $[M]_{ij} = 0$), by
noting that $[M^k(t)]_{ij}$ represents the number of walks of length $k$ that connect transaction $i$ to transaction $j$ (in a DAG, there is no difference between a walk and a path). In what follows, we assume that sites are ordered such that if transaction $i$ arrived earlier than transaction $j$, then  $i < j$. 
{\revonethree Let $t_i$ denote the time at which site $i$ was added to the DAG, i.e. the time at which it completed its PoW, $m = \max V(t)$ as the last transaction that was added to the Tangle and recall that $h$ is the duration of the PoW. Define
\be
\kappa_i = \lfloor (t_m - t_i)/h \rfloor,
\label{eq: longest path}
\ee 
\be
P_i(t) = \sum_{k=1}^{\kappa_i(t)} \mathbf{\mathbf{e_i}}^TM^k(t),
\ee 
where $\lfloor \cdot \rfloor$ is the floor operator. Then $\mathcal{H}_i(t)$ is equal to

\begin{figure}
\includegraphics[width=1\columnwidth]{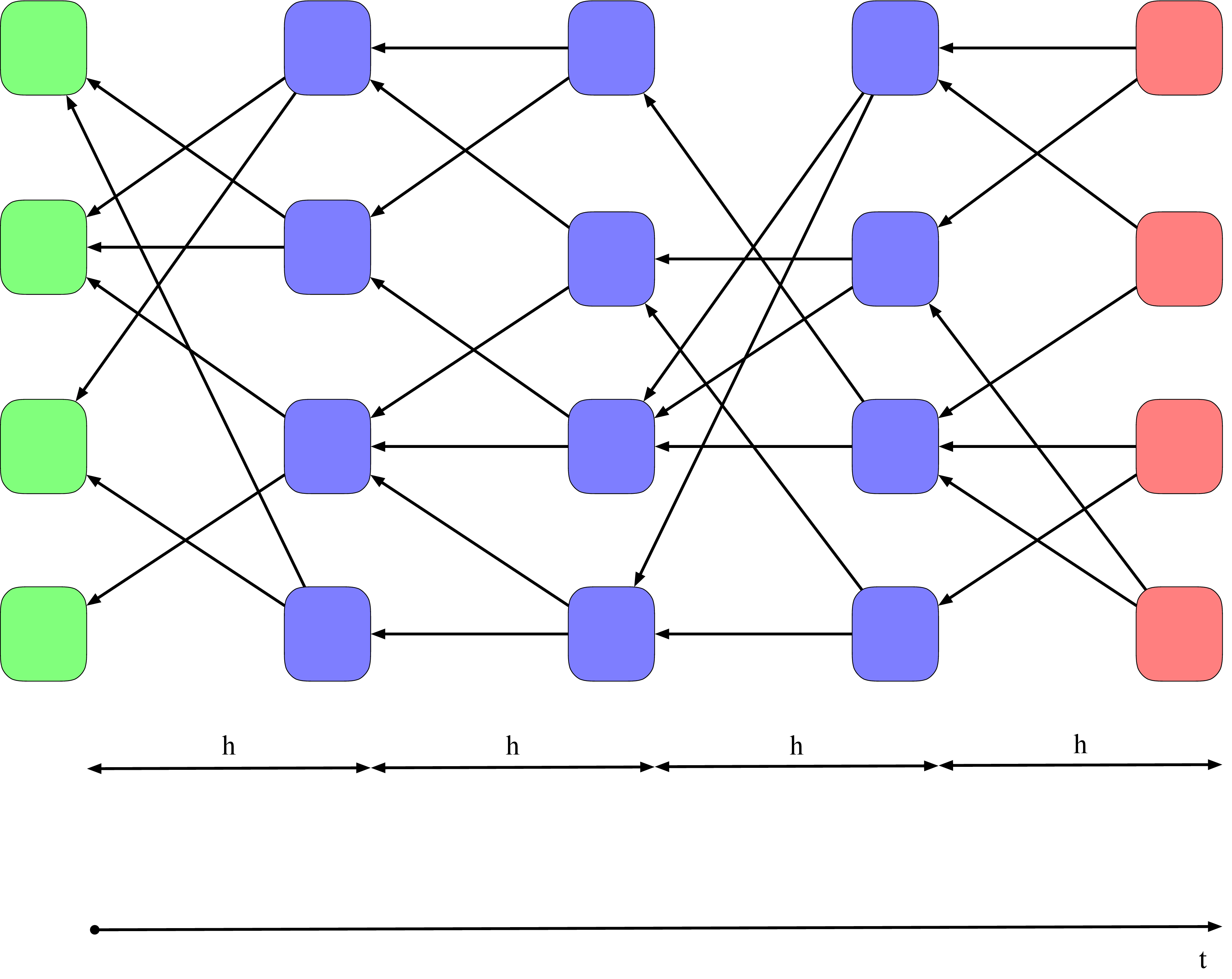}

\caption{An instance of the Tangle where the longest path between any of the leftmost transactions (highlighted in green) and the tips (highlighted in red) can not be longer than four. }
\label{Fig: Longest Path}
\end{figure}

\be
\mathcal{H}_i(t)= 1 + \sum_{j=i+1}^{N(t)} \mathrm{min}\lbrace P_i(t) \mathbf{\mathbf{e_j}},1\rbrace
\label{eq: weight}
\ee}
where $\mathbf{\mathbf{e_i}}$ is the $i$-th vector of the canonical orthonormal basis. 
Notice that:
\begin{itemize}
\item We use $\mathrm{min}\lbrace \cdot,1\rbrace$ in order to avoid counting the same transaction more than once (since  there could be several paths with a different number of steps connecting two transactions);\newline
\item {\revonethree  The value $\kappa_i$ represents the maximum number of forward steps that can occur along a directed path from site $i$ to the tips set (as each transaction must complete a PoW that lasts $h$ time units). Figure \ref{Fig: Longest Path} shows an instance of the Tangle in which the longest possible path between the green transactions and the tips set is four. Due to the presence of the PoW, a transaction whose arrival time is $t$ can have its earliest approval at time $t+h$, therefore there can be no path longer than $\kappa_i$ steps between transaction $i$ and any of the vertices that approve it (directly or indirectly).}\newline
\end{itemize}

{\revonethree To better clarify equations (\ref{eq: longest path})-(\ref{eq: weight}), consider the simple instance of the Tangle shown in Figure \ref{Fig: Cumulative Example}. The delay $h$ is set equal to 1, $m = 6$ and the adjacency matrix associated with this graph is 
\begin{figure}
\centering
\includegraphics[width=0.7\columnwidth]{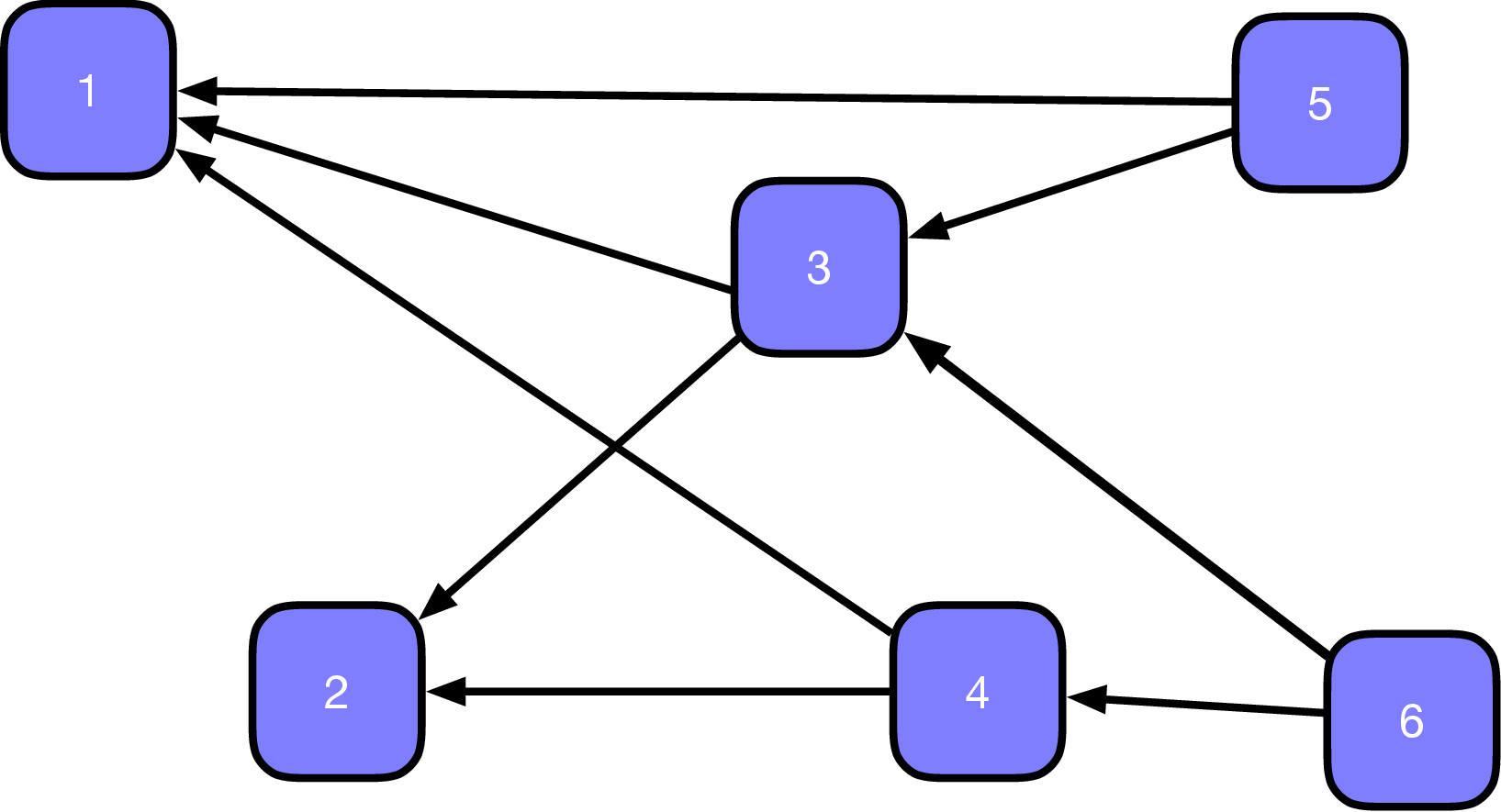}

\caption{A simple instance of the Tangle with six transactions.}
\label{Fig: Cumulative Example}
\end{figure}
\be
\nonumber
M = \begin{bmatrix}
   0 & 0 & 1 & 1 & 1 & 0 \\ 
   0 & 0 & 1 & 1 & 0 & 0 \\ 
   0 & 0 & 0 & 0 & 1 & 1 \\ 
   0 & 0 & 0 & 0 & 0 & 1 \\  
   0 & 0 & 0 & 0 & 0 & 0 \\ 
   0 & 0 & 0 & 0 & 0 & 0 \\ 
   \end{bmatrix}.
\ee
The full list of parameters can be found in Table \ref{Tab: Cumulative Weights}. Consider the first transaction: it is easy to verify by visual inspection that its cumulative weight is five. Notice, also, that $M$ is a nilpotent matrix and that $M^l = 0$ for  $l >\kappa_1$. The vector $P_1$ can be then computed as
\be
P_1 = \mathbf{\mathbf{e_1}}^T(M + M^2) = \begin{bmatrix}
   0 & 0 & 1 & 1 & 2 & 2 
   \end{bmatrix},
\ee 

and using (\ref{eq: weight}) it is easy to obtain $\mathcal{H}_1 = 5$. The same procedure can be repeated for the other transactions and the results can be found in Table \ref{Tab: Cumulative Weights}. Notice that the computations of the power of matrix $M$ need to be executed, at most, $\kappa_i$ times. \newline

\begin{table}[tb!]
\revonethree
  \caption{Tangle Parameters}
  \centering
  \begin{tabular}{ccccc}
    \hline
    $\#$ Transaction & $t_i [s]$ & $\kappa_i$ & $P_i$ & $\mathcal{H}_i$ \\
    \hline
    \hline
    1  & 0.1 & 2 & $\begin{bmatrix}
   0 & 0 & 1 & 1 & 2 & 2 
   \end{bmatrix}$ & 5 \\ 
    2  & 0.2 & 2 &  $\begin{bmatrix}
   0 & 0 & 1 & 1 & 1 & 2 
   \end{bmatrix}$ & 5 \\ 
    3  & 1.3 & 1 & $\begin{bmatrix}
   0 & 0 & 0 & 0 & 1 & 1 
   \end{bmatrix}$ & 3 \\ 
    4  & 1.6 & 1 & $\begin{bmatrix}
   0 & 0 & 0 & 0 & 0 & 1 
   \end{bmatrix}$ & 2 \\ 
    5  & 2.4 & 0 & $\begin{bmatrix}
   0 & 0 & 0 & 0 & 0 & 0 
   \end{bmatrix}$ & 1 \\ 
     6  & 2.7 & 0 & $\begin{bmatrix}
   0 & 0 & 0 & 0 & 0 & 0 
   \end{bmatrix}$ & 1 \\ 
    \hline
  \end{tabular}
  \label{Tab: Cumulative Weights}

\end{table}
}
Given (\ref{eq: weight}), we can go back to the BRW algorithm. {\revoneseven In a previous section we mentioned that in its original formulation the BRW starts at the genesis. In a practical scenario though, since the dimension of the Tangle grows with time, it would be non feasible to start the walk always from its starting transaction. A possible solution (whose effects on the exit probabilities still need to be addressed) is to start the BRW in the interior of the DAG; to take this into account we define $\mathbf{\pi}\in \mathbb{R}^{N(t)}$ as the vector whose $i$-th entry represents the probability that the walk starts at transaction $i$. To the best of the authors' knowledge no official methods have still been proposed to choose the official starting transaction. A possible solution of assigning values to $\pi$ could be based on the age and the cumulative weight of each transaction: for instance by selecting a certain interval $T=[\underline{t}, \overline{t}]$, all the transactions whose arrival time $t_i$ belongs to $T$ have a probability of being selected, as the starting point or the RW algorithm, based on their cumulative weight (i.e., the larger $\mathcal{H}_i$ the larger $\pi_i$ and vice versa). { \color{black} As another example, in \cite{LargeScale} the authors proposed to start the walk randomly, with uniform probability, at a transaction in the interval $[N(t)-200\lambda, N(t)-100\lambda]$ (where transactions are ordered by their arrival time).}}

Furthermore we define the difference in cumulative weight between transaction $j$ and transaction $k$ to be $\vartheta_{jk}(t) =\mathcal{H}_j(t)-\mathcal{H}_k(t) $. Then the transition matrix $T(t)$ whose $jk$ entry characterizes the probability of stepping from site $j$ to site $k$ is defined as follows in the case where $j$ is not a tip:

\be
[T]_{jk}(t) = \begin{cases}
	q/m & \mathrm{if\ } k \ \in \ O_j \\
	(1-q)\dfrac{e^{-\alpha\vartheta_{jk}(t)}}{\sum_{z \in I_j}e^{-\alpha\vartheta_{jz}(t)}} & \mathrm{if\ } k \ \in \ I_j \\
		0 & \mathrm{otherwise}\\
\end{cases}
\label{eq: transaction probability}
\ee
where $q\in [0,1/2)$ represents the probability of going backwards, $O_j \subset V(t)$ is the set of all sites that are directly approved by $j$, $m = | O_j |$ is the number of sites directly approved by $j$, $I_j \subset V(t)$ is the set of all sites that directly approve $j$, and $\alpha$ is a tuning parameter. When the walk hits a tip, it remains there indefinitely, therefore when the transaction $j$ is a tip we have
\be
[T]_{jk}(t) = \begin{cases}
	1 & \mathrm{if\ } j = k \\
	0 & \mathrm{otherwise}.\\
\end{cases}
\label{eq: leaf probability}
\ee
{\revonefive Note that \eqref{eq: transaction probability} appears in \cite{Equilibria}, where the BRW algorithm is analysed in a different context related to non-cooperative game theory. Here, however, we expand on this model in the interest of our attack analysis.}
Equations (\ref{eq: transaction probability}) and (\ref{eq: leaf probability}) provide us with useful information on the BRW algorithm: the walk process is an absorbing Markov Chain with $N(t) - L(t)$ transient states and $L(t)$ absorbing states, where $N(t) = |V(t)|$ is the number of sites in the tangle at time $t$,
and $L(t)$ is the number of tips at time $t$.
Therefore, by properly rearranging and re-labelling the sites the transition matrix can be written as follows:

\[ T (t) =
\begin{pmatrix}
  \begin{matrix}
  Q(t)
  \end{matrix}
  & \rvline & R(t) \\
\hline
  0 & \rvline &
  \begin{matrix}
  I
  \end{matrix}
\end{pmatrix}
\]
where $Q(t)$ is the transition matrix between transient states, $R(t)$ is the transition matrix from transient states to absorbing states, and $I$ is the identity matrix for the absorbing states. Standard analysis of absorbing Markov chains yields the absorbing probability matrix $B(t) = (I-Q(t))^{-1}R(t)$, whose $(i,k)$ entry is the probability for the BRW to be absorbed at tip $k$ given that it started at site $i$.\newline

Next we include the effects of conflicting sub-DAGs. Define $\mathcal{L}_i(t)$ to be the set containing the indices of the tips of type $i$ at time $t$, and recall that
$\mathbf{\pi}$ is the initial probability distribution of the random walk. Then the 
probability for the random walk to
terminate at a tip of type $i$ (i.e., to be absorbed by the subset $\mathcal{L}_i$ of the absorbing states) is
\be\label{def: p_i}
p_i(t) =  \dfrac{\sum_{j\in \mathcal{L}_{i}(t)}\mathbf{\pi}^TB\mathbf{\mathbf{e_j}}  }{\sum_{k=1}^d[\sum_{j\in\mathcal{L}_{k}(t)}\mathbf{\pi}^TB\mathbf{e_j}] }
\ee

Furthermore the probability for a new transaction at time $t$ to join the tip set of type $i$ is then
\be \label{eq: P_PC}
P(\mathrm{\text{Type} = i}) = p_i(t)^m \, \left(\sum_{j=1}^d p_j(t)^m \right)^{-1}
\ee
where the second factor on the right side of (\ref{eq: P_PC}) accounts for the requirement that all $m$ selections must have the same type.\newline

\begin{figure}
\includegraphics[width=1\columnwidth]{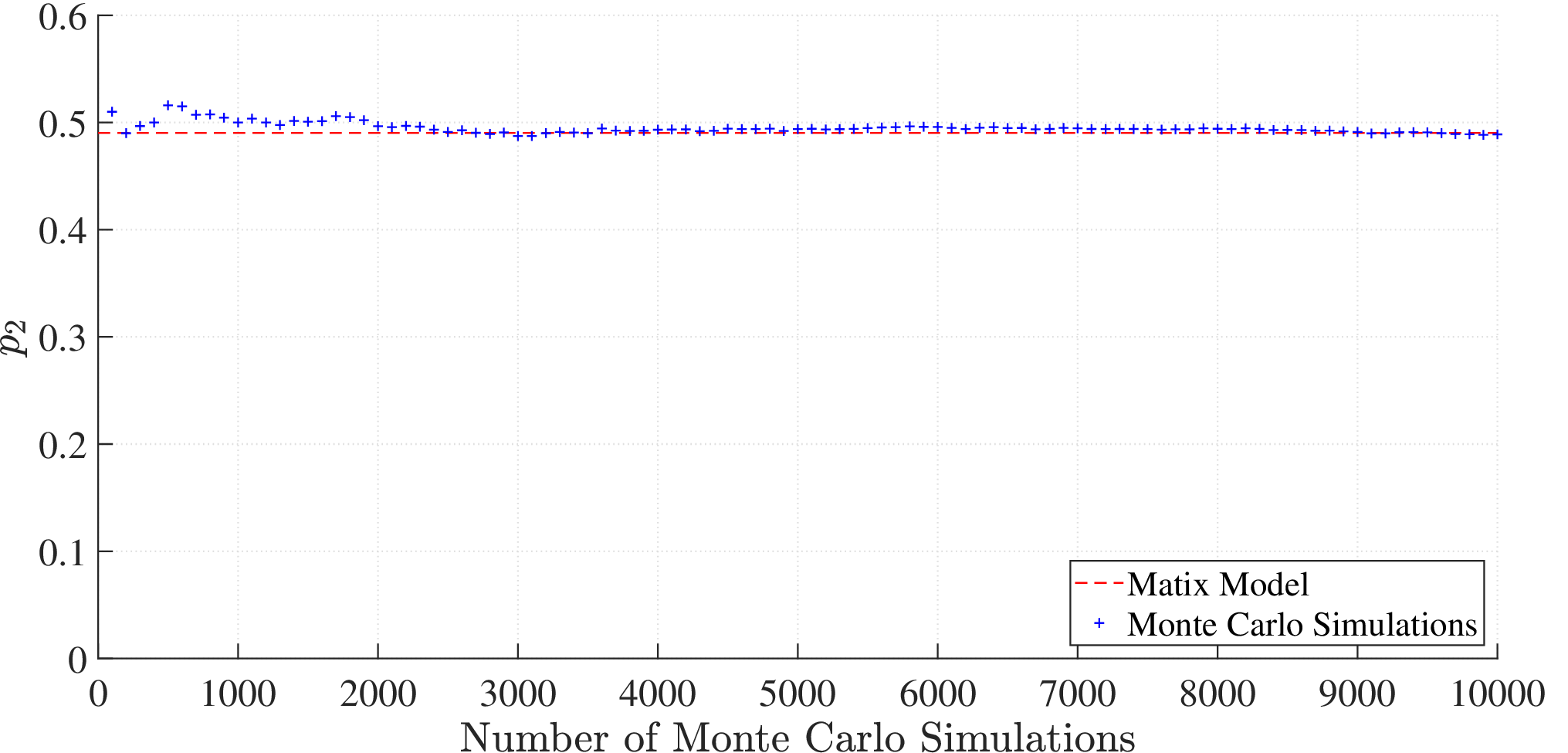}

\caption{Convergence of BRW monte carlo simulation results to matrix model formula}
\label{Fig: Convergence}
\end{figure}

{\revoneeight The validity of equation (\ref{def: p_i}) was investigated using a Monte Carlo analysis of random walks on a randomly generated Tangle with two sub-DAGs---the main DAG generated using the agent-based model, and a large parasite chain attached at a site deep in the main DAG.
In each run of the Monte Carlo analysis a random walk  was generated starting at the genesis, and the type of the terminal tip of the walk was recorded as either Type 1 or Type 2.
This is illustrated in Figure \ref{Fig: Convergence}, and we observe that as the number of Monte Carlo simulations increases the empirical probability distribution of the BRW's type converges to the value computed by \eqref{def: p_i}. Note that a very large parasite chain was attached in this example, so the value of $p_2$ is coincidentally very close to $0.5$.} \newline

\subsection{Resistance to Parasite Chain Attacks}
In this section we report on our investigations of the SPC (simple Parasite Chain) attack on the Tangle, when the BRW algorithm is used to select tips. In particular we are interested in how the parameters of the SPC and of the BRW algorithm affect the likelihood of a particle terminating on a dishonest tip. Notice that for an SPC attack to succeed, the majority of newly arriving transactions must eventually validate tips on the SPC. While it should be worthwhile to investigate the probability of this event, it would require an analysis of the whole dynamical system and of its equilibrium states, and would stretch beyond the scope of this paper. Furthermore, the probability of this event would anyway be closely related to the probability of a single BRW selecting two tips of the SPC, which is the focus of the analysis presented here. \newline

We denote tips on the main Tangle (which reference the first of the double spend transactions) as Type 1, and tips on the SPC as Type 2. Then \eqref{def: p_i} gives us:

\be
p_2(t) = \dfrac{\sum_{j\in \mathcal{L}_{2}(t)}\mathbf{\pi}^TB\mathbf{\mathbf{e_j}}  }{\sum_{l\in\mathcal{L}_{1}(t)}\mathbf{\pi}^TB\mathbf{e_l} + \sum_{j\in\mathcal{L}_{2}(t)}\mathbf{\pi}^TB\mathbf{e_j} }
\ee

\begin{figure}
\includegraphics[width=1\columnwidth]{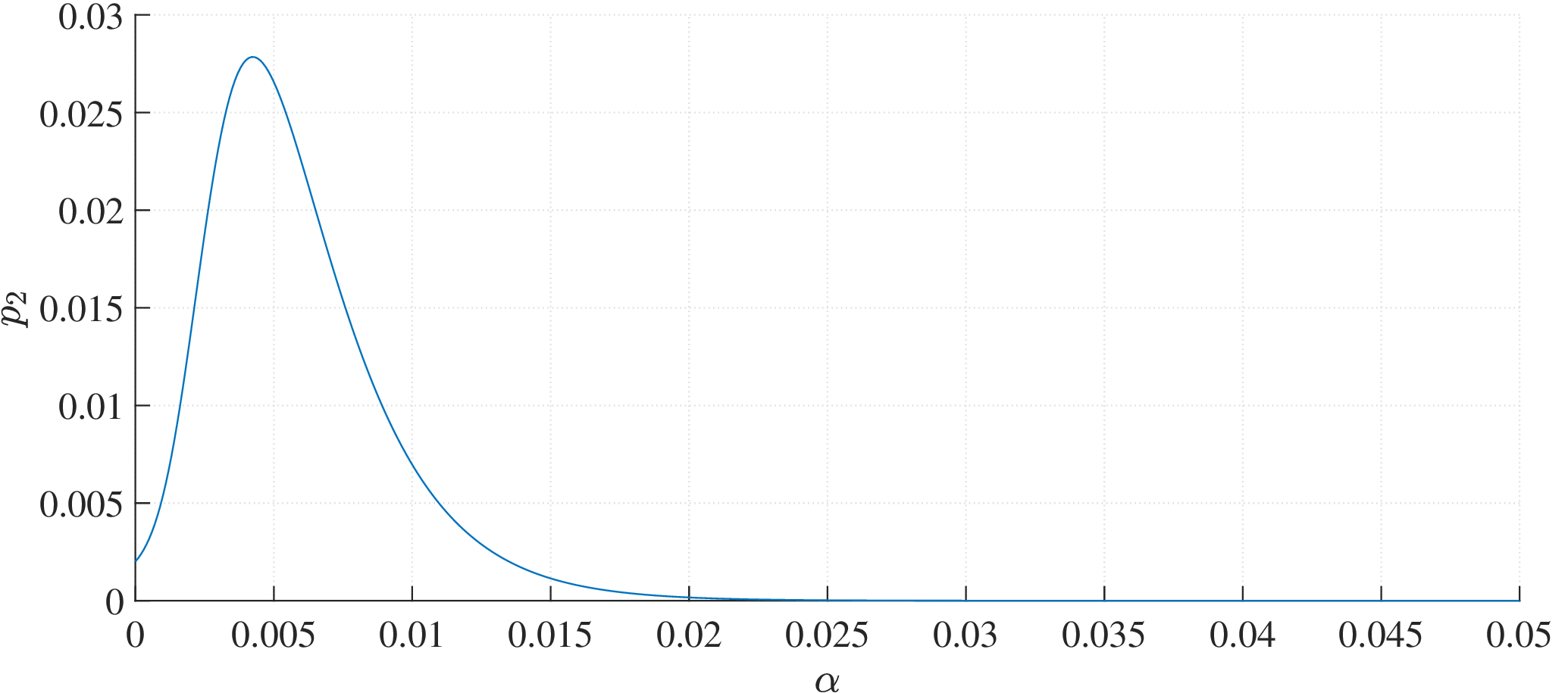}

\caption{Probability of selecting an SPC tip: $\lambda=15$, $\mu=5$, $k=1$, $T_{DS}=120$}
\label{Fig: SPC beta0}
\end{figure}

We first investigated the effect of the parameter $\alpha$ by simulating this probability for a randomly generated instance of the Tangle and a first order SPC with $T_{DS} =120$ seconds and where the attacker has 25\% of the total computing power of the network (i.e., $\mu=\lambda/3$). Results are shown in Figure \ref{Fig: SPC beta0}. {\color{black}Notice how as $\alpha$ increases, the probability of selecting the SPC increases at first (from a non zero value) reaching a maximum, and then decreases again: the reason for this behaviour is that as $\alpha$ approaches infinity the BRW will automatically select the tips with the highest cumulative weight (deterministically if only one tip has the highest cumulative weight) and, therefore, as long as $\mu<\lambda$, the attack will always fail. On the other hand, when $\alpha$ is equal to zero the selection will not depend on the cumulative weight but only on the structure of the graph. Moreover, note that, since the attacker's aim is to be able to effectively spend its money twice the attack would fail if he revealed the parasite chain before the original transaction had gained sufficient cumulative weight to be considered as confirmed by other nodes. This happens because if the original transaction was orphaned before its confirmation, then it would not be considered valid by the network and the doublespending transaction would become the only transaction in which the attacker was able to successfully spend any currency. Accordingly, in the simulations presented in Figure \ref{Fig: SPC beta0}, we assume that by $T_{DS}=120$ the initial spend on the main branch will be confirmed and, consequently, acted upon.} 

\begin{figure}
\includegraphics[width=1\columnwidth]{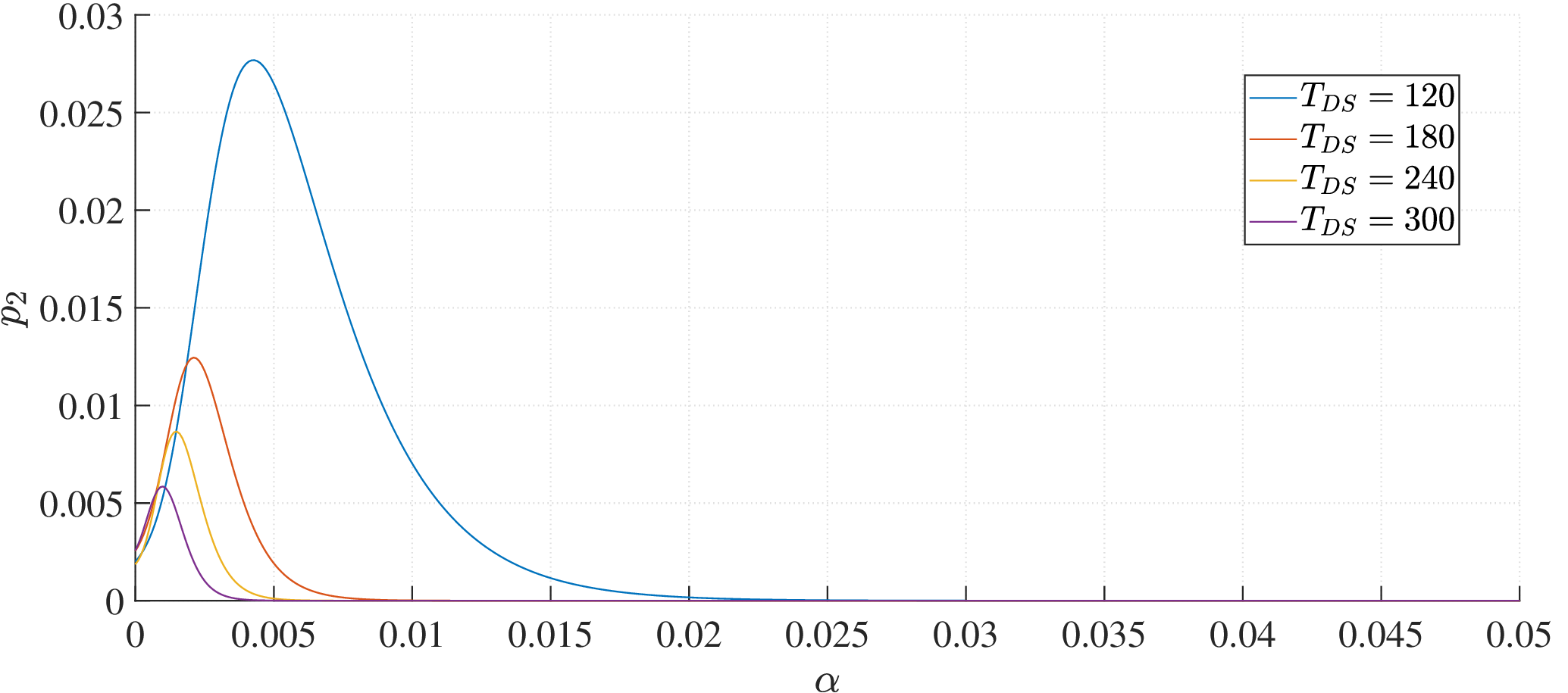}

\caption{Probability of selecting an SPC tip: $\lambda=15$, $\mu=5$, $k=1$}
\label{Fig: SPC T_DS}
\end{figure}

\begin{figure}
\includegraphics[width=1\columnwidth]{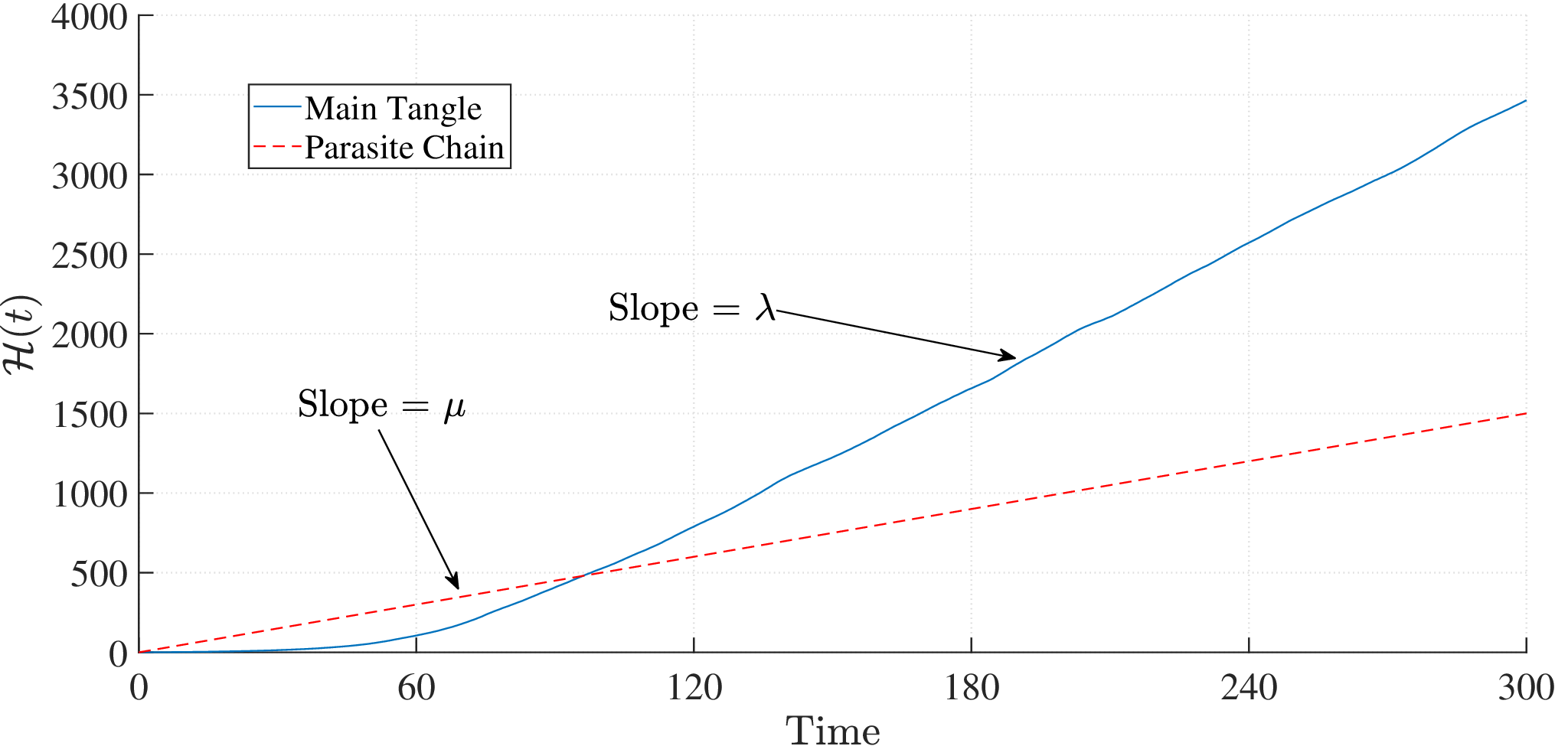}

\caption{Growth of cumulative weight in the Main Tangle and a Parasite Chain}
\label{Fig: CWG}
\end{figure}

\begin{figure}
\includegraphics[width=1\columnwidth]{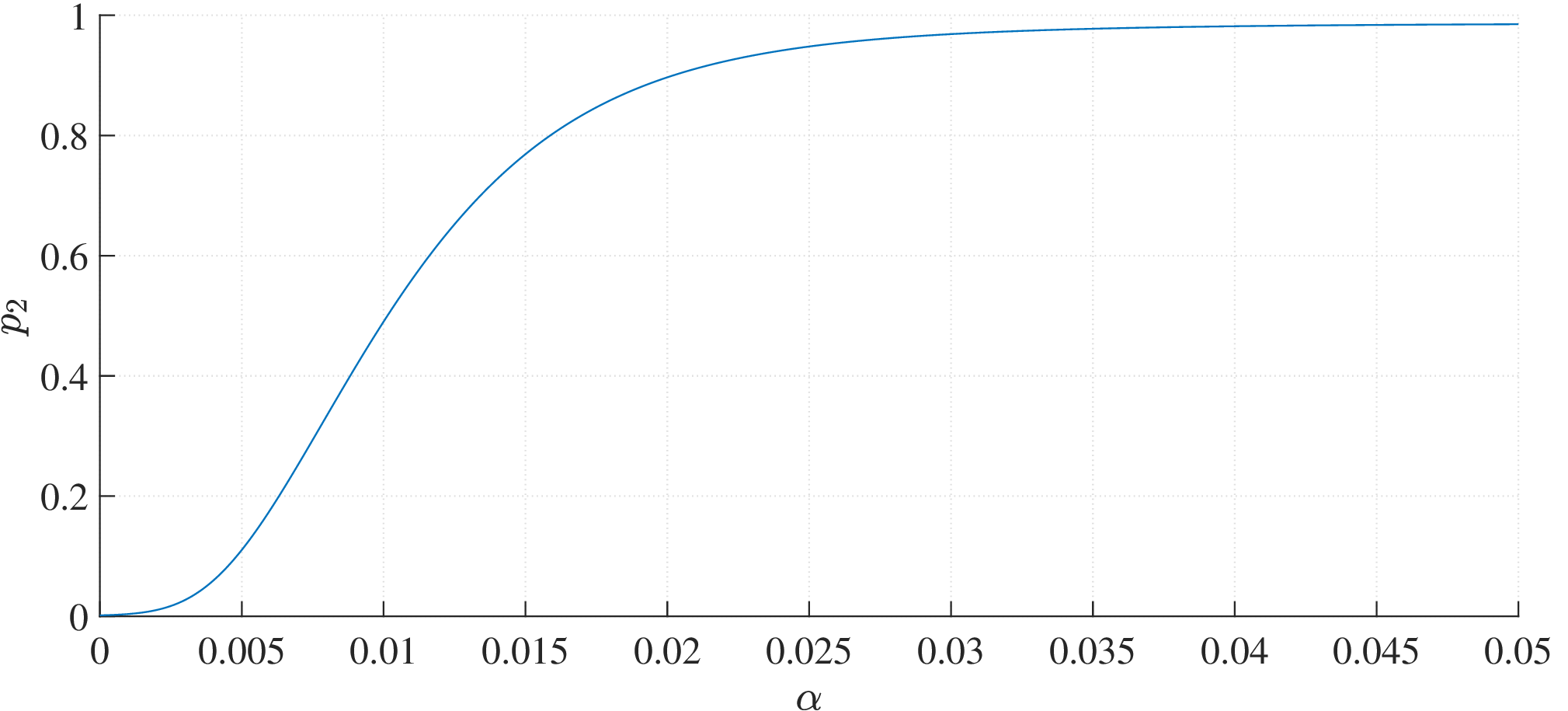}

\caption{Probability of selecting an SPC tip: $\lambda=15$, $\mu=5$, $k=1$, $T_{DS}=60$}
\label{Fig: SPC 60}
\end{figure}

Next, we investigated the effect of the double spend time, $T_{DS}$, and the results are shown in Figure \ref{Fig: SPC T_DS}. Recall that $T_{DS}$ is the time elapsed between the arrival time of the original transaction and the time the SPC was broadcast to the network. The reason for the evident decrease in probability of selecting an SPC tip as $T_{DS}$ increases, as shown in Figure \ref{Fig: SPC T_DS}, is related to how the cumulative weights of transactions grow with time, as illustrated in Figure \ref{Fig: CWG}. Because of the Tangle's structure, a site in the graph must wait for an adaptation period before all new arriving transactions will indirectly approve it, and hence increase its cumulative weight linearly. However each transaction in the SPC references directly the doublespending site, and hence its cumulative weight will immediately grow at the rate of arrival of the attacker's transactions, which is $\mu$. Figure \ref{Fig: CWG} was generated by simulating a 100 random DAGs with a parasite chains using the agent-based model. A new transaction was selected at random from the main DAG and the parasite chain during each simulation, and the trajectory of their cumulative weights were measured---Figure 12 plots the average of these 100 pairs of trajectories. The gap between the two averaged trajectories is the difference in cumulative weight, which directly reduces $p_2$ and explains the reduction in $p_2$ as we increase $T_{DS}$ that we see in Figure \ref{Fig: SPC T_DS}. The intersection point of the two curves in Figure \ref{Fig: CWG} indicates that as $T_{DS}$ decreases below this threshold, the probability of selecting an SPC tip increases (as the cumulative weight of the SPC gets larger compared to the one of the main Tangle). Figure \ref{Fig: SPC 60} confirms this hypothesis: for $T_{DS} = 60$ increasing $\alpha$ leads to an increase in the probability of selecting a tip that belongs to the doublespending attack (note that the first transaction would most likely not have been confirmed by this time so this apparent advantage may be of no use to the attacker). \newline

\begin{figure}
\includegraphics[width=1\columnwidth]{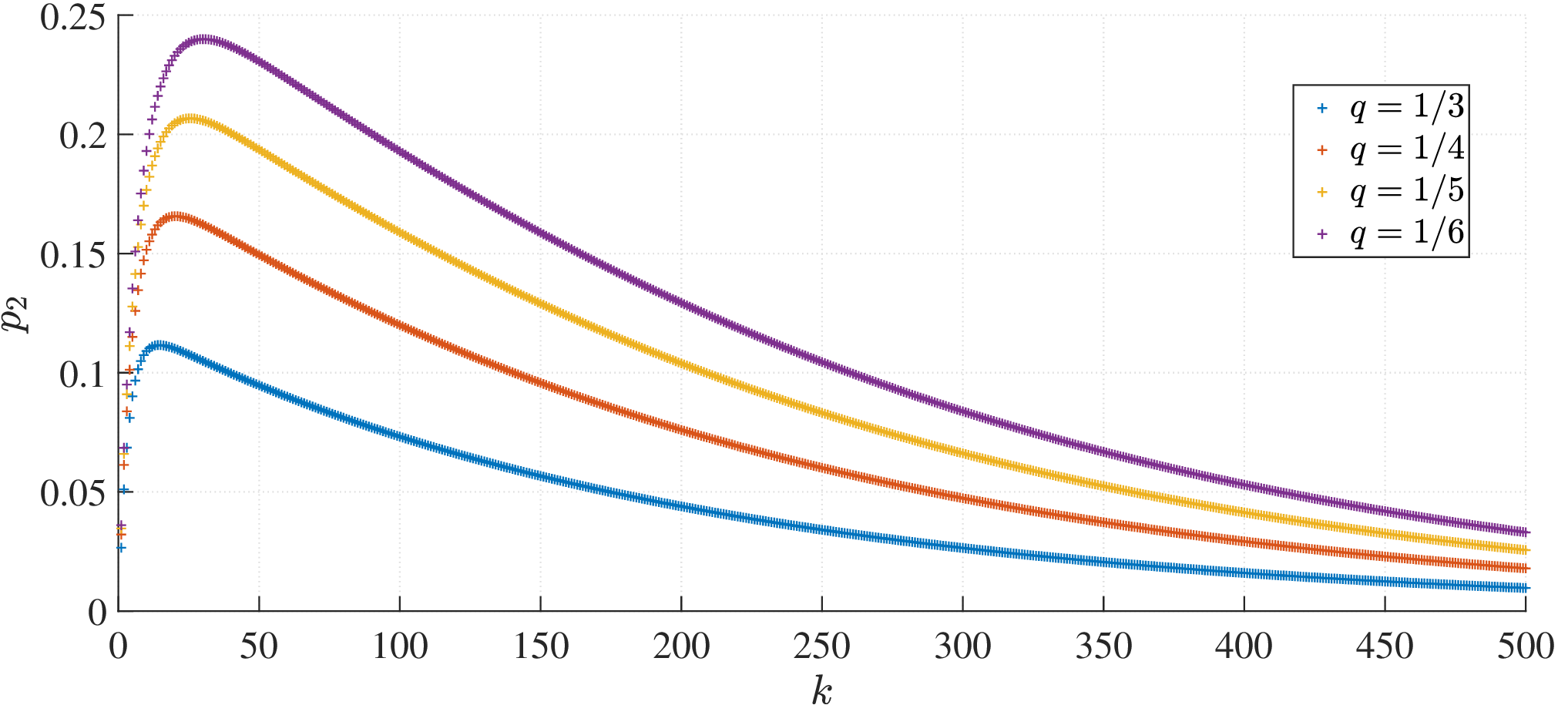}

\caption{Probability of selecting an SPC tip: $\lambda=15$, $\mu=5$, $\alpha = 0.005$, $T_{DS} = 120$} 
\label{Fig: SPC kq}
\end{figure}

The other key parameter of the SPC is the order, $k$. There are two factors to consider for an attacker when it comes to choosing $k$. The larger the choice of $k$, the more likely the BRW is to step on to the parasite chain. However, more links to the main Tangle also means more opportunities to step back on to the Tangle from the SPC (since at every step the particle has the chance to backstep). The relationship between the probability of selecting an SPC tip, the attacker's choice of $k$, and the BRW backstepping probability parameter $q$ is illustrated in Figure \ref{Fig: SPC kq}. \newline

{\color{black} As a final note, we want to stress that to maximise its chance of success, an attacker will try to attach the SPC by the time the original transaction gets confirmed by the network, independently of its computing power $\mu$. In fact, to wait less time would lead one of the two transactions to be orphaned before they get confirmed. To wait more time, on the other hand, would also reduce the attacker's chance of success, as it would lead to the original transaction’s cumulative weight to grow larger. Therefore, the parameter $T_{DS}$, while dependent on the time it takes for a transaction to be considered confirmed by the network (which may depend on various heuristics), is independent of $\mu$.}

\section{Extending the BRW algorithm} \label{sec: first order BRW}
The original motivation for using the BRW selection algorithm was to incentivise network users to validate the most recently arrived tips,  and thereby defend the Tangle against attacks such as the parasite-chain attack. Theoretically, if the BRW parameter $\alpha$ is high enough and if the attacker does not possess the majority of the computing power of the network, then the attack should never succeed. However an excessively high value of $\alpha$ would result in many honest transactions never being approved. The design trade-off between security and liveness involved in choosing $\alpha$ is discussed in \cite{POBLB}, and solutions to the issue of transactions being orphaned are proposed in \cite{TACPaper}. \newline

We now propose a modification to the BRW algorithm which seeks to reduce the efficacy of double spending attacks whilst allowing us to maintain a low $\alpha$, and hence a wide Tangle and low probability of orphaned transactions. The intuition for our modification stems from the phenomenon illustrated in Figure \ref{Fig: CWG}: although the cumulative weight of a transaction on the main Tangle may lag behind due to the initial adaptation period it underwent, we can be quite sure that if the point of attachment of a parasite chain is deep in the Tangle, then the rate of growth of the cumulative weight of transactions at the attachment site should be equal to $\lambda$. Our modification utilises the first order time derivative of the cumulative weight in calculating the stepping probabilities of the BRW---we call this a \emph{First Order} BRW. Define $\vartheta^{(1)}_{jk}(t)= |\mathcal{H}'_j(t) -\mathcal{H}'_k(t)| $, where $\mathcal{H}'_k(t)$ is the time derivative of the cumulative weight:

\begin{equation}\label{eq: extended transaction probability}
[T]_{jk}(t) = \begin{cases}
	q/m & \mathrm{if\ } k \ \in \ O_j \\
	(1-q)\dfrac{e^{-\alpha\vartheta_{jk}(t)-\beta\vartheta^{(1)}_{jk}(t)}}{\sum_{z \in I_j}e^{-\alpha\vartheta_{jz}(t)-\beta\vartheta^{(1)}_{jz}(t)}} & \mathrm{if\ } k \ \in \ I_j \\
		0 & \mathrm{otherwise}
\end{cases}
\end{equation}
where $\beta$ is a non-negative tuning parameter. Of course the time derivative $\mathcal{H}'_k(t)$ must be computed using a suitable discrete approximation (for example using the backward-difference operator and storing the previous value of the cumulative weight of each node).

The rationale for this approach can be summarized as follows: the cumulative weight of a transaction in a parasite chain grows linearly with rate equal to the computing power of the attacker, $\mu$, whilst the main Tangle will grow at the rate of the computing power of the rest of the network, $\lambda$, as illustrated in Figure \ref{Fig: CWG}. {\revtwoseven In other words, since the sites belonging to the main DAG will be growing at rate $\lambda$, and the site on the SPC will be growing with rate $\mu$, due to the \emph{honest majority} assumption ($\mu<\lambda$)  the main DAG will be favoured when the first order term is included.}  Therefore, the parasite chain will be heavily penalized by the First Order BRW. 

{\revonenine Notice that, while computing the first order derivative comes at a minimal cost (the simpler approximation requires only the previous value of the current cumulative weight) this might lead to additional hardware requirements (e.g., storage). Furthermore, it is possible that this extension could open up new attack vectors---a thorough analysis to investigate this possibility and its consequences would need to be considered before implementing this extension in a real network.} \newline

\begin{figure}
\includegraphics[width=1\columnwidth]{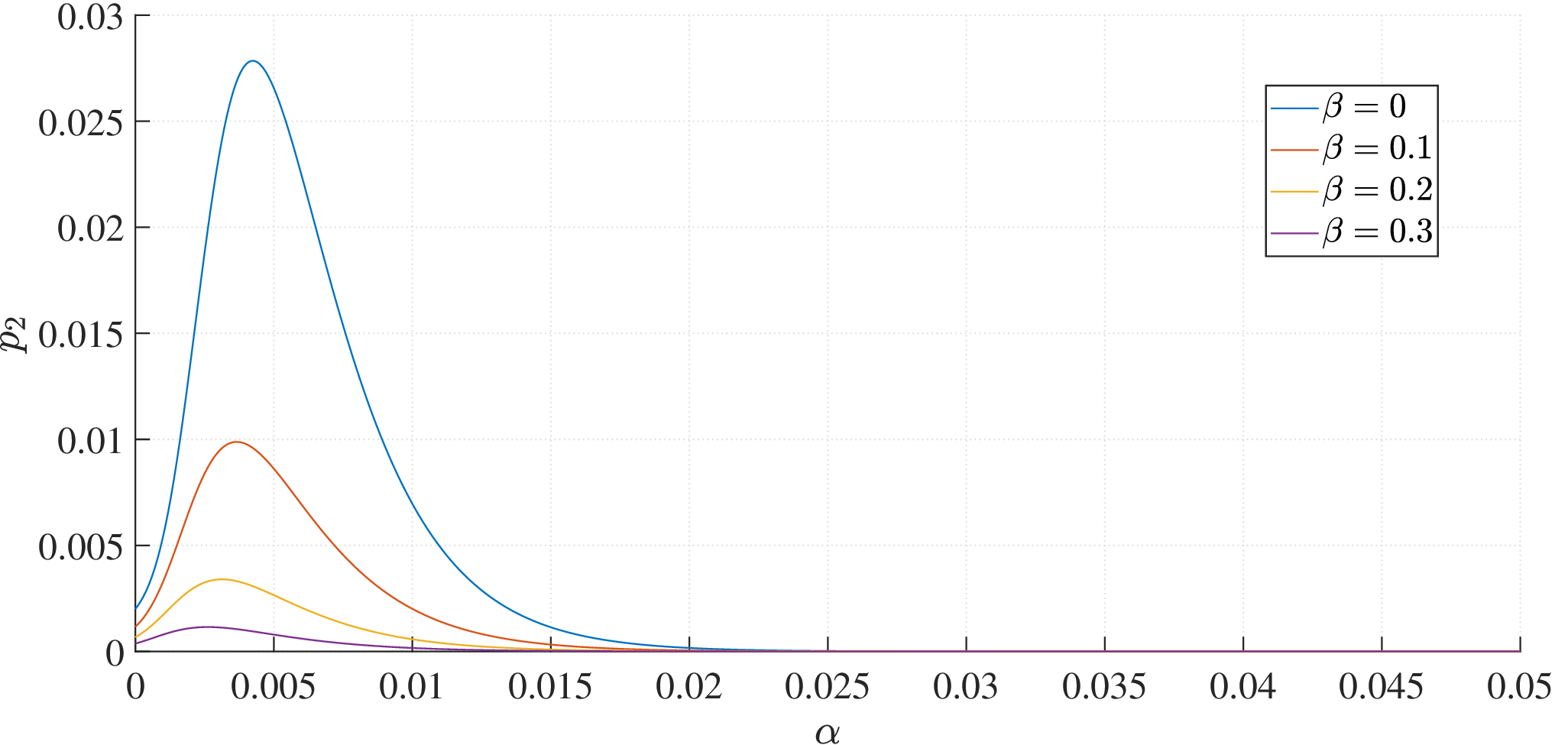}

\caption{Probability of selecting an SPC tip: $\lambda=15$, $\mu=5$, $k=1$, $T_{DS}=120$}
\label{Fig: SPC beta}
\end{figure}

Simulation results for the same instance of the Tangle used for examples in Section \ref{sec: BRW} are shown in Figure \ref{Fig: SPC beta}. These preliminary results suggest that the First Order BRW achieves its goal and effectively mitigates the chance of an attacker to achieve its goal. Of course, the performance of the algorithm will start to deteriorate when the particle approaches a tip, as at this height, the cumulative weight on the main Tangle grows at a different rate. Therefore, it could be useful to modify the derivative term adding a weight that makes it less relevant with each step forward.\newline

Finally, it is worth mentioning that (\ref{eq: extended transaction probability}) can be extended by considering higher order derivatives (in a real scenario the arrival rate of new transactions would vary in time, as would  the higher order derivatives of the cumulative weight). Future work will focus on  determining how additional terms might be added to the stepping probabilities in order to further reduce the effect of a double spending attack.

\section{Conclusions and Future Work} \label{sec: conclusions}

In this paper we considered the suitability of DAG-based DLT for applications in the IoT domain, with particular attention to a ledger's security against tampering. To this end we investigated the performance of the IOTA Tangle under a well known doublespending attack scenario called the parasite chain. Our analysis uses a Markov Chain model for the Biased Random Walk (BRW) tip selection algorithm on the IOTA Tangle (referred to as MCMC in the Tangle white paper \cite{Popov}), and we validated this model using Monte Carlo simulations of random walks on randomly generated instances of the Tangle. We then presented an extension of the BRW algorithm called \emph{First order}-BRW, which makes use the first order time derivative of the core metric of the Tangle, the cumulative weight. Our simulations demonstrated the modified algorithm's effectiveness at mitigating parasite chain attacks, compared to the standard BRW algorithm. Future lines of research will focus on analysing the effects of the use of higher order derivatives of the cumulative weight to further extend the BRW algorithm {\revonenine and to possible attack vectors that these modifications might be vulnerable to}. Furthermore, we will investigate the impact of different structures for the parasite chain attack, beyond the simple parasite chain discussed here, to assess the security properties of DAG-based ledgers.

\end{document}